\documentclass[journal]{IEEEtran}
\usepackage{cite}
\usepackage{amsmath,amssymb,amsfonts}
\usepackage{algorithmic}
\usepackage{graphicx}
\usepackage{textcomp}
\usepackage{subfigure}
\usepackage{adjustbox}
\usepackage{color}
\usepackage[backref]{hyperref}
\usepackage{multirow}
\ifCLASSINFOpdf
\else
\fi
\begin{document}
%
% paper title
% Titles are generally capitalized except for words such as a, an, and, as,
% at, but, by, for, in, nor, of, on, or, the, to and up, which are usually
% not capitalized unless they are the first or last word of the title.
% Linebreaks \\ can be used within to get better formatting as desired.
% Do not put math or special symbols in the title.
\title{GCN-MIF: Graph Convolutional Network with Multi-Information Fusion for Low-dose CT Denoising}
\author{Kecheng Chen, \IEEEmembership{Graduate Student Member, IEEE}, Jiayu Sun, Jiang Shen, Jixiang Luo, Xinyu Zhang, Xuelin Pan, Dongsheng Wu, Yue Zhao, Miguel Bento, Yazhou Ren, \IEEEmembership{Member, IEEE}, and Xiaorong Pu
\thanks{This work was supported in part by Sichuan Science and Technology Program (Nos. 2020YFS0119 and 2021YFS0172) and National Natural Science Foundation of China (No. 61806043).}
\thanks{K. Chen, J. Luo, X. Zhang, Y. Zhao, M. Bento, Y. Ren, and X. Pu are with the School of Computer Science and Engineering, University of Electronic Science and Technology of China (UESTC), Chengdu, 611731, China.}
\thanks{J. Sun and X. Pan are with the  Radiology Department, West China Hospital, Sichuan University, Chengdu, 610041, China.}
\thanks{J. Shen and D. Wu are with the  Radiology Department, West China No.4 Hospital, Sichuan University, Chengdu, 610041, China.}
\thanks{Co-corresponding authors: Xiaorong Pu and Yazhou Ren (e-mails: puxiaor@uestc.edu.cn; yazhou.ren@uestc.edu.cn;).  K. Chen and J. Sun  contributed equally to this paper.}
}

% note the % following the last \IEEEmembership and also \thanks - 
% these prevent an unwanted space from occurring between the last author name
% and the end of the author line. i.e., if you had this:
% 
% \author{....lastname \thanks{...} \thanks{...} }
%                     ^------------^------------^----Do not want these spaces!
%
% a space would be appended to the last name and could cause every name on that
% line to be shifted left slightly. This is one of those "LaTeX things". For
% instance, "\textbf{A} \textbf{B}" will typeset as "A B" not "AB". To get
% "AB" then you have to do: "\textbf{A}\textbf{B}"
% \thanks is no different in this regard, so shield the last } of each \thanks
% that ends a line with a % and do not let a space in before the next \thanks.
% Spaces after \IEEEmembership other than the last one are OK (and needed) as
% you are supposed to have spaces between the names. For what it is worth,
% this is a minor point as most people would not even notice if the said evil
% space somehow managed to creep in.

% The paper headers
\markboth{Journal of \LaTeX\ Class Files,~Vol.~14, No.~8, August~2015}%
{Shell \MakeLowercase{\textit{et al.}}: Bare Demo of IEEEtran.cls for IEEE Journals}
% The only time the second header will appear is for the odd numbered pages
% after the title page when using the twoside option.
% 
% *** Note that you probably will NOT want to include the author's ***
% *** name in the headers of peer review papers.                   ***
% You can use \ifCLASSOPTIONpeerreview for conditional compilation here if
% you desire.

% If you want to put a publisher's ID mark on the page you can do it like
% this:
%\IEEEpubid{0000--0000/00\$00.00~\copyright~2015 IEEE}
% Remember, if you use this you must call \IEEEpubidadjcol in the second
% column for its text to clear the IEEEpubid mark.

% use for special paper notices
%\IEEEspecialpapernotice{(Invited Paper)}

% make the title area
\maketitle

% As a general rule, do not put math, special symbols or citations
% in the abstract or keywords.
\begin{abstract}
Being low-level radiation exposure and less harmful to health, low-dose computed tomography (LDCT) has been widely adopted in the early screening of lung cancer and COVID-19. LDCT images inevitably suffer from the degradation problem caused by complex noises. It was reported that deep learning (DL)-based LDCT denoising methods using convolutional neural network (CNN) achieved impressive denoising performance. Although most existing DL-based methods (e.g., encoder-decoder framework) can implicitly utilize non-local and contextual information via downsampling operator and 3D CNN, the explicit  multi-information (i.e., local, non-local, and contextual)
integration  may not be explored enough. To address this issue, we propose a novel \underline{g}raph \underline{c}onvolutional \underline{n}etwork-based LDCT denoising model, namely GCN-MIF, to explicitly perform \underline{m}ulti-\underline{i}nformation \underline{f}usion for denoising purpose. Concretely, by constructing intra- and inter-slice graph, the graph convolutional network is introduced to leverage the non-local and contextual relationships among pixels. The traditional CNN is adopted for the extraction of local information. Finally, the proposed GCN-MIF model fuses all the extracted local, non-local, and contextual information. Extensive experiments show the effectiveness of our proposed GCN-MIF model by quantitative and visualized results. Furthermore, a double-blind
reader study on a public clinical dataset is also performed to validate the usability of denoising results in terms of the structural fidelity, the noise suppression, and the overall score. Models and code  are available at \href{https://github.com/tonyckc/GCN-MIF\_demo}{https://github.com/tonyckc/GCN-MIF\_demo}.
\end{abstract}


\section{Introduction}
\IEEEPARstart{C}{omputed} tomography (CT) is one of the most frequently used imaging technologies in  modern medicine \cite{buzug2011computed}.  Compared with conventional radiography, CT has the advantages of superior contrast resolution \cite{lischer2005fracture}, superb detailed anatomical representations \cite{kim2007antibiofouling} and the ability to selectively enhance or remove structures from images \cite{Rosso2007Preliminary}.
However, recent studies report that the radiation exposure of CT scans may come with potential cancer risks, especially for  children \cite{PEARCE2012499}.  In the past two decades, low-dose computed tomography (LDCT) thus has become a hot screening tool for noninvasive low radiation examination, such as the early detection of lung cancer \cite{PEARCE2012499} and  the diagnosis of COVID-19 pneumonia \cite{tabatabaei2020low}. Compared with normal-dose CT (NDCT), The reduction of dose resulting in heavily noisy CT images is a though challenge, since it will affect the diagnostic accuracy for the radiologists. 
%\begin{figure*}
%    \centering
%   \centerline{\includegraphics[width = \textwidth]{figs/image1.pdf}}
%    \caption{The workflow of radiologists when reading the LDCT images. In step 1, the radiologist focuses on the region of interest (ROI) with local information. In step 2, the radiologist leverages those non-local but easily-observed similar tissues to perform auxiliary observation. In step 3,  the radiologist slides the mouse so that the similar ROI in the front and rear slices can be observed.}
%    \label{fig1}
%\end{figure*}

%To overcome this problem, quite a few studies aim to %remove the noise from acquired LDCT images,  in order to % obtain the latent noise-free CT images
To overcome this problem, quite a few studies make an attempt to obtain the latent noise-free CT images by removing noise from LDCT images
\cite{RN314,liu2012adaptive,8340157,choi2020statnet}. Existing LDCT denoising methods can be roughly divided into three streams. The first two streams are \textit{sinogram filtration} \cite{wang2005sinogram,manduca2009projection,balda2012ray} and \textit{iterative reconstruction}  \cite{liu2012adaptive,zhang2016tensor} based methods, respectively. Both of them achieve an effective denoising performance on account of involving the projection data directly. It is also an intractable problem that the acquisition of  projection data is fairly complicated in clinical environments \cite{8340157}. In addition, the \textit{iterative reconstruction}
based methods need to transform the data  from the projection  domain to image domain constantly \cite{10.1007/978-3-030-63830-6_36}, which is usually regarded as a time consuming process \cite{8964295}.  
Thanks to the progress of deep neural network, the \textit{post-processing} based methods using  deep neural networks (DNN) \cite{10.1007/978-3-030-63830-6_36,RN314,RN308,choi2020statnet} realize excellent denoising performance compared with the first two streams.  Instead of relying heavily on projection data, deep learning-based LDCT denoising methods work in the image domain of CT data directly \cite{chen2017low},  which is extremely convenient. The time consumption of deep learning-based methods is significantly lower than  the first two streams \cite{RN314}.
\begin{figure*}
    \centering
    \centerline{\includegraphics[width = \textwidth]{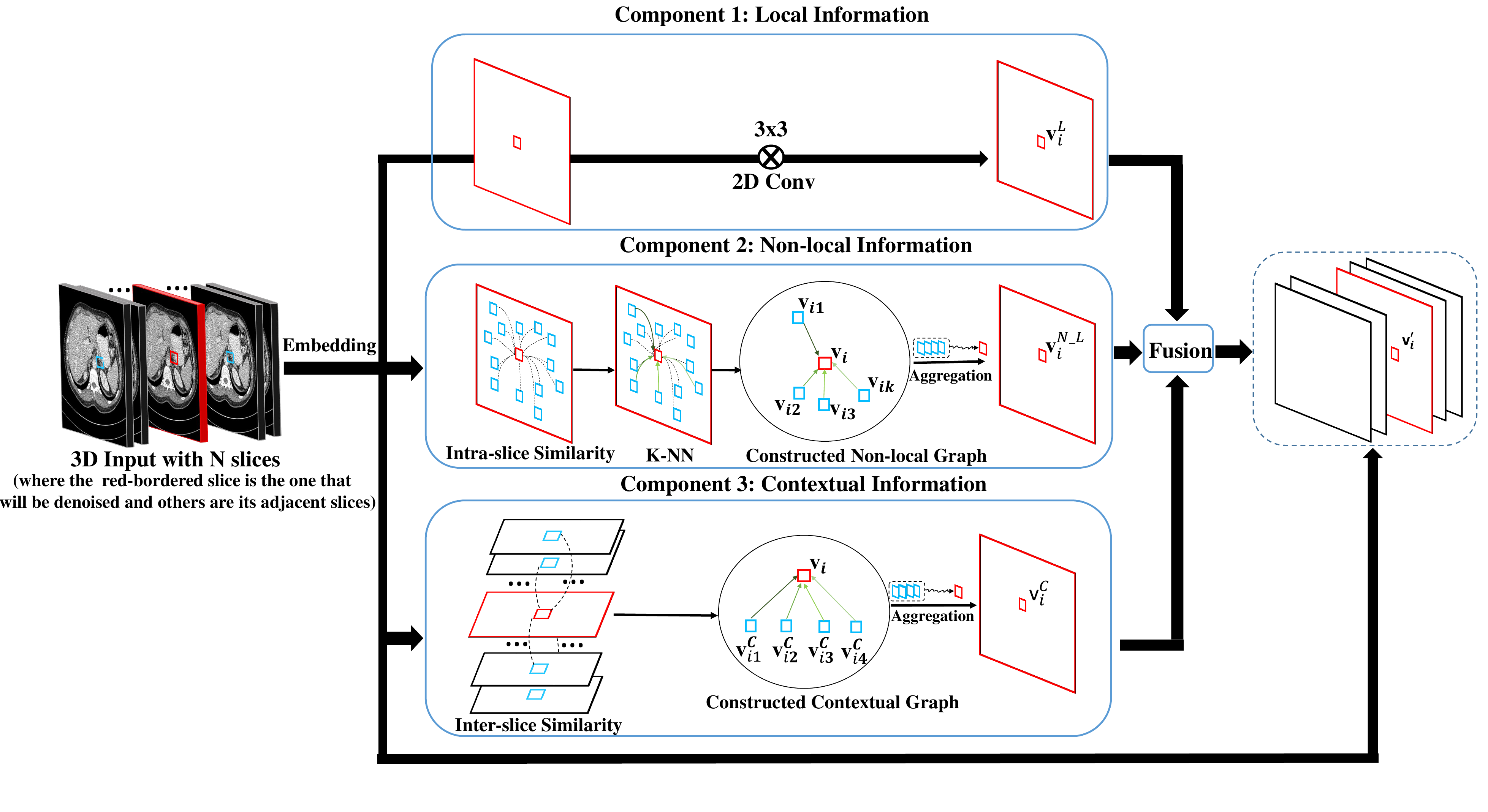}}
    \caption{The proposed graph convolutional network with multi-information fusion (GCN-MIF) module. Followed by extraction of the feature map, the classical 2D convolution operation is applied in component 1 and the computation of the intra-slice similarity, the computation of a K nearest neighbor (K-NN), the construction of the non-local graph, as well as the feature aggregation is successively performed in component 2. In component 3, the whole feature maps are inputted in and then compute the inter-slice, the construction of the context graph and the feature aggregation. Finally, the result of feature fusion is used to replace the feature map in the original input position, in order to keep the original 3D shape \cite{8964295}. Please see the section \hyperref[GCN]{MATERIALS AND METHODS} for more details.}
\end{figure*}

By constructing a parameterized denoising model (a.k.a., denoiser), most existing deep learning-based LDCT denoising methods usually learn a supervised mapping \cite{chen2017low,chen2017Low2} or texture transfer \cite{yang2018low,10.1007/978-3-030-63830-6_36,8340157} from LDCT images to corresponding NDCT ones. Specially, convolutional neural network (CNN)-based denoisers with encoder-decoder framework \cite{yang2018low,chen2017Low2,RN308,fan2019quadratic} are widely adopted due to their impressive performance. For example, Residual Encoder-Decoder Convolutional Neural Network \cite{chen2017Low2} (RED-CNN) and Conveying Path-based Convolutional Encoder-decoder \cite{RN308} (CPCE) proposed to stack symmetrical downsampling (i.e. convolution operator) and upsampling (i.e. deconvolution operator) operations for denoising purpose. Recently, some researches \cite{fan2019quadratic,zhang2021artifact,zhang2021clear} proposed to concentrate on the improvement of elabrated structurual details based on baselined CNN-based models. For instance, Zhang et al.\cite{zhang2021artifact} proposed a multi-channel denoiser to maintain the detail information of denoising LDCT images. A subtle structure enhanced LDCT denoising model \cite{zhang2021clear} was also proposed. More recently, unsupervised LDCT denoising methods \cite{gu2021cyclegan,kwon2021cycle} were proposed to address the lack of paired LDCT/NDCT images, where cycle Generative Adversarial Network (cycleGAN) \cite{zhu2017unpaired} is a popular unpaired LDCT image denoising framework \cite{gu2021adain,kang2019cycle}. For example, Kwon et al. \cite{kwon2021cycle} proposed to impose a cycle-free cycleGAN for unpaired LDCT image denoising. One can notice that existing cycleGAN-based methods typically adopt the CNN-based model as the generator. 

As discussed above, the design of the denoising model still plays a crucial role in supervised (as the denoiser) and unsupervised (as the generator) LDCT image denoising methods. Although existing CNN-based models achieved very impressive denoising performance \cite{zhang2021clear,gu2021cyclegan,kwon2021cycle,chen2017Low2}, the convolution operator essentially is a local operator that extracts some useful local information within it's filter size \cite{li2020sacnn}. Meanwhile, 3D CNN-based encoder-decoder model (e.g., \textit{UNet} model \cite{li2018h}) can extract non-local and contextual information with the increasing of receptive field. However, as commented by \cite{wang2020non}, stacking downsampling convolutional layers is an implicit manner to aggregate long-range (a.k.a., \textit{non-local}) information, which may cause
the loss of more spatial information during encoding compared with the explicit one. Thus, some works \cite{wang2020non} aimed to introduce more explicit non-local module into Unet model (e.g., \textit{non-local Unet} \cite{wang2020non}) for medical image segmentation. 

For LDCT image denoising tasks, the self-attention module \cite{shaw2018self} is designed to explicitly exploit the non-local relationships between one pixel with all the pixels. Compared with encoder-decoder model \cite{chen2017Low2,RN308}, some obvious improvements, especially for the suppression of the noise and the preserving of structural information,  were reported by the self-attention-based model (e.g., SACNN) \cite{li2020sacnn}. However, the introduction of too many non-local relationships may be not optimal, because 1) the denoising performance may be gradually saturated with the increasing of the number of non-local relationships, and 2) it is a huge inference and computational consumption to utilize all pixel-by-pixel relationships on the overall feature map. 
%according to the experimental results. In contrast, GCN module can exploit optimal non-local neighbors flexibly. Computationally efficient
%Although these CNN-based methods achieve impressive performance, the explicit non-local and contextual information (which are also leveraged by the radiologist as mentioned in aforementioned discussions)  are typically ignored by most existing methods.  The most important issue that should be noted is, radiologists may not support a model whose workflow is vastly different from theirs. 
\begin{figure*}

    \centering
    \centerline{\includegraphics[width = \textwidth]{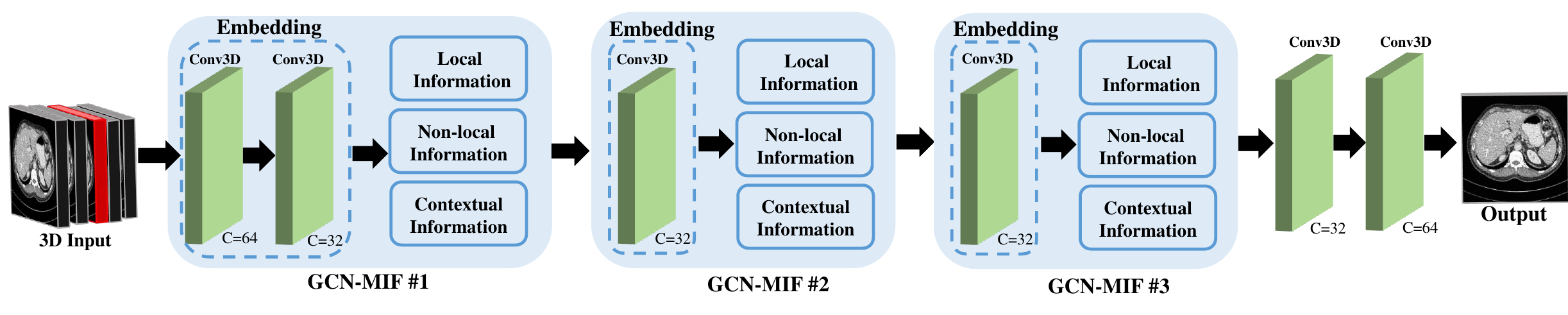}}
    \caption{The proposed stacked GCN-MIFs framework. This model stacks three GCN-MIFs modules. The last two convolution layers aim to generate a channel encoder-decoder structure as suggested in \cite{8964295,RN308}.}
\end{figure*}

To this end, we propose a novel graph convolutional network-based LDCT denoising model, namely GCN-MIF, to explicitly leverage the fused representation of local, non-local, and contextual information (i.e. multi-information) for denoising purpose. 
Specifically, instead of using the non-local relationships between one pixel with all the pixels (e.g., self-attention), we propose to construct the intra-slice graph through top-$K$ similarity among pixels (as the non-local neighbors), which can enjoy the advantages of \textit{flexible usage of non-local relationships} and \textit{high computational efficiency}. Then, the graph convolutional network (GCN) is introduced to aggregate those obtained non-local relationships, which can be regarded as a more explicit manner of extracting non-local information compared with CNN-based models. Furthermore, by establishing inter-slice graphs, the GCN is proposed to utilize contextual dependencies among pixels. Compared with the contextual information implicitly extracted by 3D CNN \cite{RN308,li2020sacnn}, the information extracted by GCN is also explicit. Here, the GCN modules for intra- and inter-slice information are proposed to construct a novel deep learning module, namely 3D GCN module, which aims to consider the non-local and contextual information together. Obviously, the local information is also important. We therefore leverage classical convolution operation to acquire useful local patterns. Finally, the proposed GCN-MIF integrates the aforementioned three components (i.e., the local, non-local, and contexutal information) to learn the best adaptive composition through the feature fusion. Note that the proposed GCN-MIF can also be stacked to further improve the denoising capacity. In this paper, the adopted model is constructed by three GCN-MIFs due to the consideration of comprehensive performance, as shown in Figure 3. 
%On the other hand, existing deep learning-based LDCT denoising methods can be roughly divided into two categories, i.e., mean square error (MSE)-guided CNN methods and generative adversarial network (GAN)-based methods. 

\section{Materials and methods}
\subsection{PRELIMINARY} \label{GCN} First, some preliminaries need to be claimed. Given that an LDCT-NDCT paired dataset, $T$=$\{(\mathbf{x}_{1},\mathbf{y}_{1}),(\mathbf{x}_{2},\mathbf{y}_{2}), \cdots, (\mathbf{x}_{N},\mathbf{y}_{N})\}$, where \textit{N} denotes the number of paired training samples. $(\mathbf{x},\mathbf{y}) \in (\mathcal{X},\mathcal{Y})$, where $\mathcal{X}$ and $\mathcal{Y}$ are two image domains, respectively. The paired samples $\mathbf{x}$ $\in$ $\mathbb{R}^{n \times n}$ and $\mathbf{y}$ $\in$ $\mathbb{R}^{n \times n}$ are matrix expression of a noisy LDCT image and a high-quality NDCT image. Assume that the $i$th slice $\mathbf{x}_{i}$ is the one that needs to be denoised in overall slices of a patient. The 3D input of $\mathbf{x}_{i}$ can be represented as $\mathbf{X}_{i} = concat(\mathbf{x}_{-s},\mathbf{x}_{i-1},\mathbf{x}_{i},\mathbf{x}_{i+1},\cdots,\mathbf{x}_{s})$, where $2s+1$ is a predefined number of slices and set to 3 as suggested in \cite{li2020sacnn}. The $concat$ operator denotes the concatenation operator along with the first dimension. 
The goal of the LDCT image denoising process can be represented as below,
\begin{equation}
    \hat{\mathbf{y}} = F_{\phi}(\mathbf{X}),
\end{equation}
where $\hat{\mathbf{y}}$ denotes the estimated high-quality NDCT image (as the denoising result). The parametrized network $F_{\phi}$ aims to learn an appropriate mapping that can minimize the difference between $\hat{\mathbf{y}}$ and $\mathbf{y}$. In the framework of GAN, 
$F_{\phi}$ also can be represented as a generator, which aims to produce the plausible sample. The design of the denoising network thus is extremely important to determine the denoising result. To this end, a novel graph convolutional network with multi-information fusion (namely GCN-MIF) is proposed in this paper. More details will be presented as following.

\subsection{The methodology of proposed GCN-MIF model}
The proposed GCN-MIF model consists of four components, i.e., the layer of embedding, the layer of extracting local information, the 3D graph convolutional networks (3D GCN), and the layer of feature fusion.  We will discuss them in details. 

\subsubsection{The layer of embedding} It is usually more effective for model learning to convert the input into its embedding in feature space \cite{valsesia2019deep}. Every pixel then can be regarded as the format of the feature vector along with the channel. To this end, the classical 3D convolution operation is adopted in our proposed GCN-MIF model. Specifically, the first GCN-MIF model has two 3D convolution layers with channels of 64 and 32, due to the need of stronger capacity of feature extraction in the start of a model. Other GCN-MIF models only have one 3D convolution layer with channel of 32. To maintain the size of the feature map, the operation of ‘reflect’ padding is applied, which is useful to avoid the artifacts of the edge as suggested in \cite{valsesia2019deep}. The output of embedding layer thus can be denoted as $\mathbf{X'}_{i} =(\mathbf{x'}_{-s},\mathbf{x'}_{i-1},\mathbf{x'}_{i},\mathbf{x'}_{i+1},\cdots,\mathbf{x'}_{s})= embedding(\mathbf{X}_{i})$.

\subsubsection{The 3D graph convolutional networks}  As the most important component in the proposed GCN-MIF model, the 3D graph convolutional networks (namely 3D GCN) consists of two modules, i.e., plane graph convolutional module and depth graph convolutional module, respectively.  

\textbf{Plane Graph Convolutional Module.} For every pixel in the feature space of a CT slice, we propose to construct a plane graph (a.k.a., intra-slice graph), $G_{p} = (V,E)$. Note that we only extract non-local information of the feature map that needs to be denoised, i.e., the $\mathbf{x'}_{i}$ of $\mathbf{X'}_{i}$, due to the considerations of efficiency and the contribution to the final result. Assume that there are $K$ vertices  in a plane graph $G_{p}$, represented as $v_{i} \in V$ for a vertex. Each edge is a pair of vertices, represented as $(v_{i}, v_{j}) \in E$, ${\forall} i < j$.  Our goal is to represent the non-local relations among features and aggregate those non-local information. 

The first step is to indicate the importance of the vertex $j$'s feature to vertex $i$'s one. In image domain, the importance of one pixel relatively to another is usually measured by similarity \cite{li2020sacnn}.  Instead of relying on the weight matrix to obtain the importance, e.g., self-attention \cite{he2016deep},  in this paper, we propose to express the importance by the pixel-wise Euclidean distance in the feature space, which is inspired by  Non-local Means (NLM)
\cite{10.1109/CVPR.2005.38} for patch-wise  non-local denoising. As a non-deep-learning method, NLM-based LDCT denoising method \cite{li2014adaptive} has impressive adaptive denoising ability for
complex noise level in real environments, which will also be desired for existing deep learning-based models. Formally, the similarity can be computed as follows,
\begin{equation}
    e_{ij} = \frac{\| v_{i} - v_{j}\|_{2}^{2}}{h_{ij}},
\end{equation}
 %This usually leads to having more stable gradient. 
where $h_{ij}$ denotes the square root of the dimension of the $v_{i}$ or $v_{j}$. In order to reduce computational complexity, we compute the $e_{ij}$ for pixels $j \in \mathcal{N}_{i}$, where $\mathcal{N}_{i}$ is a $d \times d -8$ (The total number of directly adjacent pixels for vertex $i $ is 8) non-local neighborhood region of the vertex $i$. It is worthy noting that the proposed plane graph convolutional module enjoys some potential advantages compared with self-attention-based LDCT denoising model (e.g., SACANN \cite{li2020sacnn}):
\begin{itemize}
    \item Self-attention module can be regarded to utilize all non-local neighbors according to the wights computed by inner product. However, the effect of non-local information may be gradually saturated and even decreases slightly (we will discuss this in details in the Experiment section), with the increasing of number of selected non-local neighbors. Thus, the plane graph convolutional module can flexibly select optimal non-local neighbors (in saturation point), which may contribute to better denoising performance.
    \item Due to usage of the relationships of all the pixels,
    self-attention module inevitably takes more inference time and computational consumption (Details will be presented in the Experiment section). Instead, GCN typically uses appropriate non-local information, leading to competitive performance.
\end{itemize}

 We then propose to normalize them across all choices of $j$ such that  the importance between vertices can be represented easily. 
\begin{equation}
    a_{ij}=softmax_{j}(e_{ij})=\frac{exp(\frac{-\| v_{i} - v_{j}\|_{2}^{2}}{h_{ij}})}{\sum\limits_{k=1}^{K-1}exp(-e_{ik})}.
\end{equation}
One can observe the importance of neighbor vertices $v_{j}$'s of $v_{i}$
is represented by the probabilistic result $a_{ij}$ induced from the distance of feature space. In this paper, we assign every edge $(v_{i},v_{j})$ as $a_{ij}$. The next step is to aggregate those non-local information on this weighted graph. In order to achieve the specific-task fashion, inspired by the Edge-Conditioned Convolution (ECC) \cite{simonovsky2017dynamic}, the probability-based edge is introduced into ECC as follows: 

\begin{equation}
\begin{aligned}
        \mathbf{s}_{i}^{N\_L} & = \frac{1}{K-1}\sum \limits_{j=1}^{K-1}(F^{l}(a_{ij}, w)\mathbf{v}_{ij} + b) \\
          & = \sum \limits_{j=1}^{K-1}\frac{\Theta_{ij}\mathbf{v}_{ij}}{(K-1)}+b,
\end{aligned}
\end{equation}
where $F^{l}$ denotes the output of a network parameterized by $w^{l}$ which is used to  dynamically produce the weight matrix $\Theta^{l}_{j,i}$ for different edge labels ${z}^{l,j\rightarrow i}$, and $\mathbf{b}^{l}$ is a learnable bias.  The optimization details can be found in \cite{simonovsky2017dynamic}. In our work, we adopt a multi-layer perception network for $F^{l}$, 

\textbf{Depth Graph Convolutional Module.} Although widely-adopted 3D CNN can implicitly impose contextual information, depth graph convolutional module aims to explicitly obtain the useful contextual information in the inter-slices.  For every pixel in the feature space of a CT slice, a depth graph (a.k.a., inter-slice graph) $G_{d} = (V,E)$ is constructed. Assume that there are $M$ vertices in a depth graph $G_{d}$, represented as $v_{i} \in V$ for a vertex. Each edge is a pair of vertices, represented as $(v_{i}, v_{j}) \in E$, ${\forall} i < j$.  Our goal is also to represent the context relations among features and aggregate them. We adopt the same aggregation method as described in \textit{Plane Graph Convolutional Module}. In this paper, $M$ is set to 3, as the number of slices for a 3D  input is not large.

\subsubsection{The layer of extracting local information.} In practice, the local information is extremely important for denoising. As in previous studies \cite{10.1007/978-3-030-63830-6_36}, the 2D convolution operation with the $3 \times 3$ filter is utilized to extract the local information of the feature map that needs to be denoised. Therefore, the 2D convolution operation is also introduced to explore the meaningful local information, as shown in component 1 of Figure 1.

\subsubsection{The layer of feature fusion} Until now, the non-local information, local information and contextual information have been obtained. There is a troublesome issue, i.e., how do we fuse them? Before introducing our adopted fusion method, we first analyze the significance of different information for denoising purpose. For the contextual information, if the thickness of a set of slices is very large, the relationship of inter-slice will not be very strong for learning. The difference of slice thickness for different body regions or imaging vendors may cause a similar risk. Therefore, the usability of contextual information may be not stable. Instead, the inner information (including non-local information and local information) of a slice is extremely important. To this end, a simple yet effective multi-information fusion method is utilized as following
%In order to address this issue, we must firstly review the workflow of radiologists. Intuitively, the radiologist, compared with the contextual information, will pay more attention to  Furthermore, due to the difference of slice thickness for different body regions or imaging vendors, the usability of inter-slice information may be not stable, because if the thickness of a set of slices is very large, the relationship of inter-slice will not be very strong for learning. We thus consider the contextual information into an  auxiliary information. Based on aforementioned discussions, the process of feature fusion can be 
%represented as
\begin{equation}
    \mathbf{x''}_{i} = \alpha \cdot Mean(\mathbf{p'}_{i,NL}+\mathbf{p'}_{i,L}) + (1-\alpha) \cdot \mathbf{p'}_{i,C},
\end{equation}
where $\mathbf{p'}_{i,NL}$, $\mathbf{p'}_{i,L}$, and $\mathbf{p'}_{i,C}$
denote the non-local information of $\mathbf{x'}_{i}$, the local information of $\mathbf{x'}_{i}$, and the contextual information of $\mathbf{x'}_{i}$, respectively. $\alpha \in [0,1]$ is a learnable parameter.  The $\alpha$ is initialized to $0$. $Mean$ operator denotes the pixel-wise average operation. It should be noted that  the contextual information is considered as auxiliary information based on aforementioned discussions.

Finally, the result of fusion $\mathbf{x''}_{i}$ is proposed to replace the feature map in original input position (as shown in Figure 2), i.e., $\mathbf{X'}_{i} = concat(\mathbf{x}_{-s},\mathbf{x}_{i-1},\mathbf{x''}_{i},\mathbf{x}_{i+1},\cdots,\mathbf{x}_{s})$, due to the considerations of keeping original 3D shape and a similar shortcut idea \cite{8964295}. As shown in Figure 3, the overall model stacks three GCN-MIF modules. The last two convolution layers aim to generate a channel encoder-decoder structure as suggested in \cite{8964295,RN308}.

\subsubsection{The overall framework and loss function} The proposed GCN-MIF model can be flexibly inserted into mean square error (MSE)-guided model and WGAN-VGG \cite{RN308}-based framework. Specifically, the proposed GCN-MIF model plays the role of generator $G$. The discriminator  $D$ follows the structure in the WGAN-VGG \cite{RN308}. The loss function of generator is composed of adversarial loss and perceptual loss as  \cite{RN308}
\begin{equation}
    \min_{\theta_{G}} = -\mathbb{E}_{\mathbf{X}_{i}} D(G(\mathbf{X}_{i}))+ \lambda \cdot \mathbb{E}_{\mathbf{X}_{i},\mathbf{y_{i}}}\|\phi(G(\mathbf{X}_{i}))-\phi(\mathbf{y_{i}}) \|_{2}^{2}.
\end{equation}
As in previous studies, VGG19 \cite{simonyan2014very} is adopted as the feature extractor $\phi$. $\lambda$ is a balance term that is set to 0.1 as suggested in \cite{RN308}. $\theta_{G}$ denotes the parameters of the generator. The loss function of the discriminator follows the wasserstein generative adversarial network based on gradient penalty \cite{gulrajani2017improved}. For the MSE-guided model, the
proposed GCN-MIF model is the denoiser $G$. The loss function can be represented as
\begin{equation}
    \min_{\theta_{G}} = \frac{1}{N}\|G(\mathbf{X})-\mathbf{y}\|_{F}^{2}, 
\end{equation}
 where $\|\cdot\|_{F}$ denotes the Frobenius norm. $N$ denotes the number of samples in a batch.

\begin{figure*}[!h]
  \centering
  \subfigure[]{

  \begin{minipage}{0.98\columnwidth}
  \centering
     \centerline{\includegraphics[width= 0.99\columnwidth]{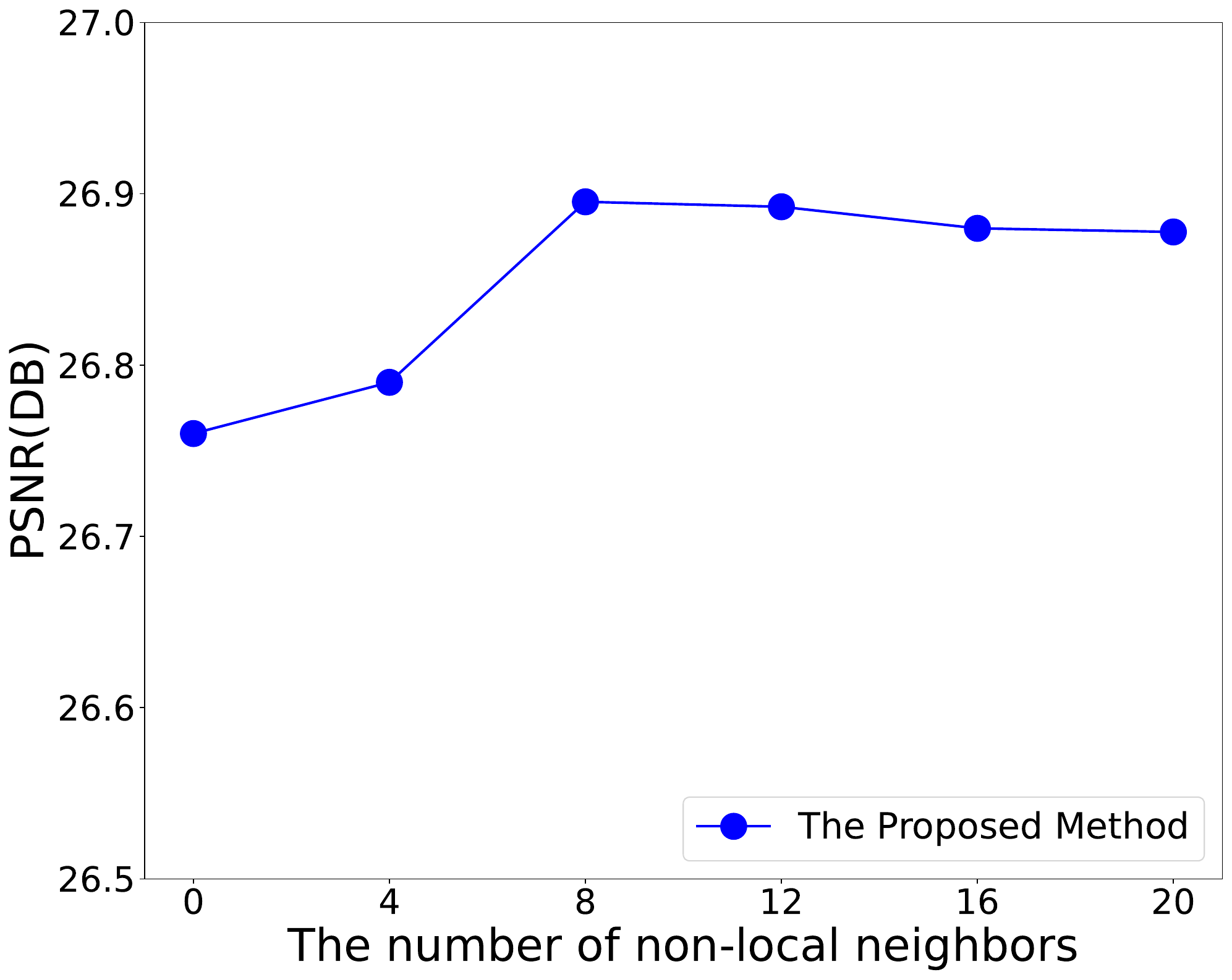}}
  \end{minipage}
  }
  \subfigure[]{

  \begin{minipage}{0.98\columnwidth}
  \centering

     \centerline{\includegraphics[width= 0.99\columnwidth]{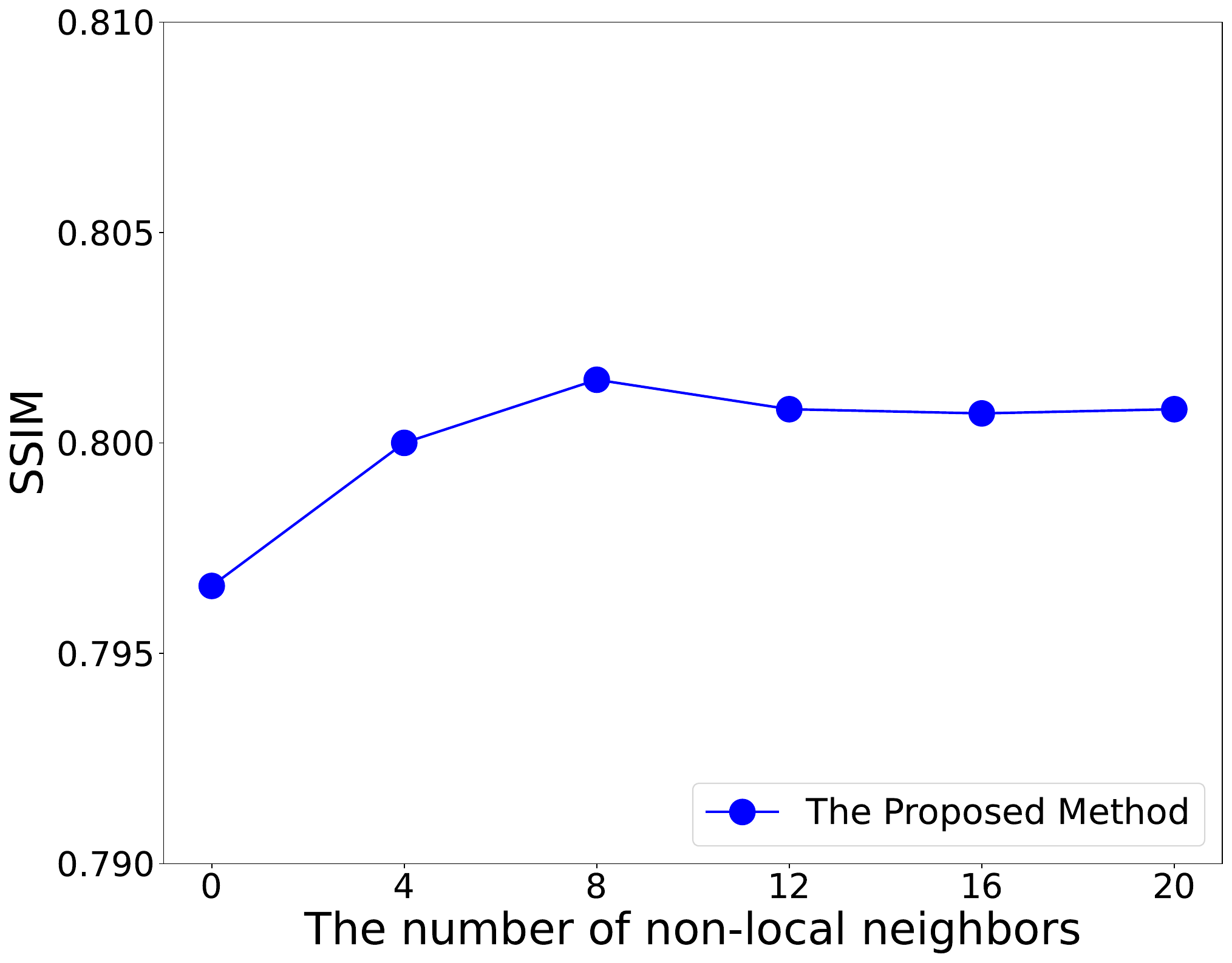}}
  \end{minipage}
  }
  \caption{The impact of the number of non-local neighbors for GCN. (a) The PSNR performance of denoising results. (b) The SSIM performance of denoising results.}
  \label{parameter modification}
\end{figure*}
\begin{figure*}[!h]
    \centering
    \centerline{\includegraphics[width = \textwidth]{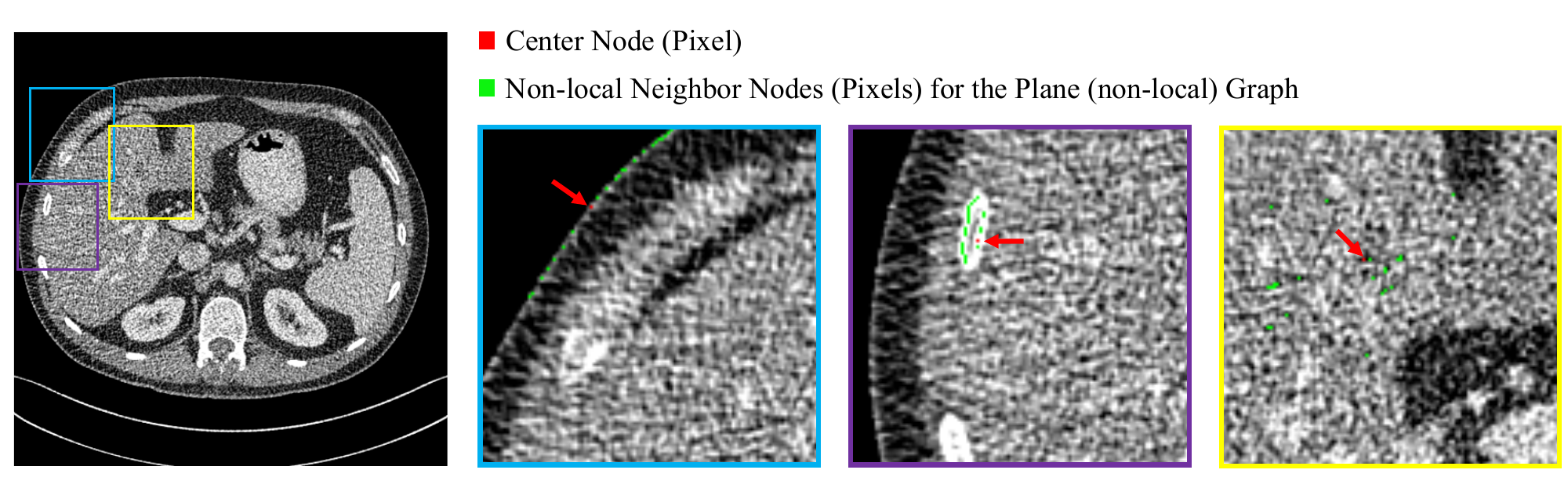}}
    \caption{The visualized results of constructed plane (non-local) graph for some pixels. Please zoom in for better view. The display window is [-160, 240] HU. }
\end{figure*}
\section{Experiments}
In this section, we first validate the effectiveness of each component of the proposed GCN-MIF model. Second, we compare the performance of different non-local modules by quantitative and visualized denoising results. Then, we compare our proposed method with baseline methods. Finally, a double-blind reader study on a public clinical dataset is performed.
\subsection{Datasets}
As previous studies \cite{RN314,8340157,choi2020statnet}, the public low/normal-dose dataset released from the 2016 NIH-AAPM-Mayo Clinic Low Dose CT Grand Challenge is used for training. This dataset includes 10 patients' abdominal examinations obtained on similar scanner models (Somatom Definition AS+, or Somatom Definition Flash operated in single-source mode, Siemens Healthcare, Forchheim, Germany). The normal-dose CT images are acquired under the settings of 120kV, 50mAs. Note that the low-dose counterparts are simulated to reach a noise level that corresponded to $25\%$ by inserting the Poisson noise. More information about the dataset is available at https://www.aapm.org/GrandChallenge/LowDoseCT/\#.  We use 6 patients' scan as the training set, 1 patient's scan as the validation set, and 3 patients' scan as the test set (where 1 patient for ablation study and 2 patients for baseline comparison). To balance the learning efficiency and the memory consumption, every CT image with the size of 512$\times$512 is randomly divided into non-overlapping 64$\times$64 sub-patches. The total number of  sub-patches is about 27K.

\subsection{Training Details.}
For the details of preprocessing of training set, we normalize the data in CT domain (the provided dataset is the format of DICOM) into 0 to 1 using different window width and window level, according to different body regions. For the abdomen scans, the window width and window level are 400, 40, respectively. For the chest scans, the  window width and window level are 1500, -600, respectively. The outputs of the model are re-normalized into corresponding range of CT domain to generate the DICOM files. For the details of model training, the tensorflow \footnote{www.tensorflow.org} is applied to construct the proposed GCN-MIF model. The Adam optimizer \cite{zhang2018improved} is used to optimize the parameters of the model with the learning rate of exponential decay. For the frameworks of generative adversarial network \cite{yi2019generative}, the initial learning rate is set to $1 \times 10^{-4}$ for the generator and $4 \times 10^{-4}$ for the discriminator as suggested in \cite{Rosso2007Preliminary}. For the framework of MSE-guided network, the learning rate is set to $1 \times 10^{-4}$. To balance the memory consumption and learning efficiency, the batch size is set to 32. The models are trained with 40 epochs and converges finally. We chose the model that has the best performance on validation set for comparison experiment.

\subsection{The impact of the number of non-local neighbors for GCN }
In the Introduction section, we argue that the usage of too many non-local relationships may be not optimal due to the denoising performance and computational consumption. Here, we set the number $K$ of non-local neighbors with different values to observe the change of denoising performance. We evaluate the peak-to-noise ratio (PSNR) and structural similarity index (SSIM) on the validation set. 

We can observe from Figure 3 that the denoising performance gradually improves with the increasing of the number of non-local neighbors. After a saturation point ($K=8$), the performance of the model decreases slightly and becomes stable finally. Therefore, it is extremely crucial to utilize appropriate non-local information, which can avoid potential negative impact and ineffective computation (we will discuss this later). The number of $K$ thus is set to 8 in this paper.

To validate the effectiveness of constructed plane (non-local) graph, we visualize selected non-local neighbors for some pixels.
As shown in Figure 4, we can find that those non-local neighbors (green point) indeed have strong relationships with their center node (red point). Interestingly, the selection of non-local neighbors even can obey anatomical structure (along with the edge), as illustrated in the first sub-figure of Figure 4.
 
\begin{table*}[!h] \label{component_comparison}
\centering
\caption{The Effectiveness of Each Component. For the PSNR and the
SSIM, the higher the better. In the MSE-guided models, the best score is bolded with the red. In the WGAN-VGG-based models, the best score is bolded with the blue. }
\renewcommand{\arraystretch}{1.4}
\begin{tabular}{|c|lcc|c|ll|ll|}
\hline
\multirow{2}{*}{Method}    & \multicolumn{3}{c|}{Self-Similarity}                                                                     & \multicolumn{1}{l|}{\multirow{2}{*}{Loss Function}} & \multicolumn{2}{c|}{PSNR}                                           & \multicolumn{2}{c|}{SSIM}                                          \\ \cline{2-4} \cline{6-9} 
                           & \multicolumn{1}{l|}{Local}               & \multicolumn{1}{l|}{Non-local} & \multicolumn{1}{l|}{Context} & \multicolumn{1}{l|}{}                               & \multicolumn{1}{c|}{MEAN}             & \multicolumn{1}{c|}{STD}    & \multicolumn{1}{c|}{MEAN}            & \multicolumn{1}{c|}{STD}    \\ \cline{1-1} \cline{5-9} 
LDCT         & \multicolumn{1}{l|}{}                    & \multicolumn{1}{l|}{}          & \multicolumn{1}{l|}{}        & -                                                   & \multicolumn{1}{l|}{23.4417}          & 1.8799                      & \multicolumn{1}{l|}{0.8075}          & 0.0478                      \\ \hline \hline
\multirow{2}{*}{Baseline1} & \multicolumn{1}{c}{\multirow{2}{*}{\checkmark CNN}} & \multirow{2}{*}{-}             & \multirow{2}{*}{\checkmark CNN}         & MSE                                                 & \multicolumn{1}{c|}{28.1532}          & \multicolumn{1}{c|}{1.5250} & \multicolumn{1}{c|}{0.8459}          & \multicolumn{1}{c|}{0.0396} \\ \cline{5-9} 
                           & \multicolumn{1}{c}{}                     &                                &                              & VGG                                                 & \multicolumn{1}{c|}{26.8347}          & \multicolumn{1}{c|}{1.5235} & \multicolumn{1}{c|}{0.8443}          & \multicolumn{1}{c|}{0.0390} \\ \hline \hline
\multirow{2}{*}{Baseline2} & \multirow{2}{*}{\checkmark CNN}                     & \multirow{2}{*}{-}             & \multirow{2}{*}{\checkmark GCN}         & MSE                                                 & \multicolumn{1}{l|}{28.2802}          & 1.5338                      & \multicolumn{1}{l|}{0.8495}          & 0.0394                      \\ \cline{5-9} 
                           &                                          &                                &                              & VGG                                                 & \multicolumn{1}{l|}{26.8693}          & 1.4544                      & \multicolumn{1}{l|}{0.8443}          & \multicolumn{1}{r|}{0.0389} \\ \hline \hline
\multirow{2}{*}{Baseline3} & \multirow{2}{*}{\checkmark CNN}                     & \multirow{2}{*}{\checkmark GCN}           & \multirow{2}{*}{\checkmark GCN}         & MSE                                                 & \multicolumn{1}{l|}{\textcolor{red}{28.3742}} & 1.5450                      & \multicolumn{1}{l|}{\textcolor{red}{0.8510}} & 0.0395                      \\ \cline{5-9} 
                           &                                          &                                &                              & VGG                                                 & \multicolumn{1}{l|}{\textcolor{blue}{27.0925}}          & 1.4986                      & \multicolumn{1}{l|}{\textcolor{blue}{0.8469}}          & \multicolumn{1}{r|}{0.0388} \\ \hline
\end{tabular}
\end{table*}
\begin{table*}[]
\caption{The comparison of different non-local models on the test set. For the PSNR and the
SSIM, the higher the better. In the MSE-guided models, the best score is bolded with the red. In the WGAN-VGG-based models, the best score is bolded with the blue. }
\renewcommand{\arraystretch}{1.6}
\begin{adjustbox}{max width=\textwidth}
\begin{tabular}{|ccccccc|ll|ll|}
\hline
\multicolumn{1}{|c|}{\multirow{2}{*}{\textbf{Method}}}    & \multicolumn{1}{c|}{\multirow{2}{*}{\textbf{Non-local Type}}} & \multicolumn{1}{c|}{\multirow{2}{*}{\textbf{Details}}}         & \multicolumn{1}{l|}{\multirow{2}{*}{\textbf{Parameters}}} & \multicolumn{1}{c|}{\multirow{2}{*}{\textbf{\begin{tabular}[c]{@{}c@{}}Inference Time\\    (in Seconds/1K Iters.)\end{tabular}}}} & \multicolumn{1}{c|}{\multirow{2}{*}{\textbf{\begin{tabular}[c]{@{}c@{}}GPU Consumption\\ (MiB)\end{tabular}}}} & \multicolumn{1}{l|}{\multirow{2}{*}{\textbf{Loss Function}}} & \multicolumn{2}{c|}{\textbf{PSNR}}                                        & \multicolumn{2}{c|}{\textbf{SSIM}}                                       \\ \cline{8-11} 
\multicolumn{1}{|c|}{}                                    & \multicolumn{1}{c|}{}                                         & \multicolumn{1}{c|}{}                                          & \multicolumn{1}{l|}{}                                     & \multicolumn{1}{c|}{}                                                                                                             & \multicolumn{1}{c|}{}                                                                                          & \multicolumn{1}{l|}{}                                        & \multicolumn{1}{c|}{\textbf{MEAN}}    & \multicolumn{1}{c|}{\textbf{STD}} & \multicolumn{1}{c|}{\textbf{MEAN}}   & \multicolumn{1}{c|}{\textbf{STD}} \\ \hline
\multicolumn{7}{|c|}{LDCT}                                                                                                                                                                                                                                                                                                                                                                                                                                                                                                                                   & \multicolumn{1}{l|}{23.4417}          & 1.8799                            & \multicolumn{1}{l|}{0.8075}          & 0.0478                            \\ \hline
\multicolumn{1}{|c|}{\multirow{2}{*}{Baseline4}} & \multicolumn{1}{c|}{\multirow{2}{*}{Encoder-Decoder}}         & \multicolumn{1}{l|}{\multirow{2}{*}{3 "Downsampling" Modules}} & \multicolumn{1}{c|}{\multirow{2}{*}{$\sim \#$0.62e5}}        & \multicolumn{1}{c|}{\multirow{2}{*}{$\sim$50}}                                                                                    & \multicolumn{1}{c|}{\multirow{2}{*}{$\sim$7.4K}}                                                               & MSE                                                          & \multicolumn{1}{c|}{27.9534}          & \multicolumn{1}{c|}{1.4898}       & \multicolumn{1}{c|}{0.8456}          & \multicolumn{1}{c|}{0.0394}       \\ \cline{7-11} 
\multicolumn{1}{|c|}{}                                    & \multicolumn{1}{c|}{}                                         & \multicolumn{1}{l|}{}                                          & \multicolumn{1}{c|}{}                                     & \multicolumn{1}{c|}{}                                                                                                             & \multicolumn{1}{c|}{}                                                                                          & VGG                                                          & \multicolumn{1}{c|}{26.1236}          & \multicolumn{1}{c|}{1.3846}       & \multicolumn{1}{c|}{0.8413}          & \multicolumn{1}{c|}{0.0330}       \\ \hline
\multicolumn{1}{|c|}{\multirow{2}{*}{Baseline5}} & \multicolumn{1}{c|}{\multirow{2}{*}{Self-Attention}}          & \multicolumn{1}{c|}{\multirow{2}{*}{3 Self-Attention Modules}} & \multicolumn{1}{c|}{\multirow{2}{*}{$\sim \#$2.7e5}}         & \multicolumn{1}{c|}{\multirow{2}{*}{$\sim$750.80}}                                                                                & \multicolumn{1}{c|}{\multirow{2}{*}{$\sim$18K}}                                                                & MSE                                                          & \multicolumn{1}{l|}{28.3304}          & 1.5313                            & \multicolumn{1}{l|}{0.8498}          & 0.0401                            \\ \cline{7-11} 
\multicolumn{1}{|c|}{}                                    & \multicolumn{1}{c|}{}                                         & \multicolumn{1}{c|}{}                                          & \multicolumn{1}{c|}{}                                     & \multicolumn{1}{c|}{}                                                                                                             & \multicolumn{1}{c|}{}                                                                                          & VGG                                                          & \multicolumn{1}{l|}{26.4418}          & 1.4269                            & \multicolumn{1}{l|}{0.8422}          & \multicolumn{1}{r|}{0.0392}       \\ \hline
\multicolumn{1}{|c|}{\multirow{2}{*}{Ours}}      & \multicolumn{1}{c|}{\multirow{2}{*}{GCN}}                     & \multicolumn{1}{c|}{\multirow{2}{*}{3 "GCN" Modules}}        & \multicolumn{1}{c|}{\multirow{2}{*}{$\sim \#$2.8e5}}         & \multicolumn{1}{c|}{\multirow{2}{*}{$\sim$360}}                                                                                   & \multicolumn{1}{c|}{\multirow{2}{*}{$\sim$9K}}                                                                 & MSE                                                          & \multicolumn{1}{l|}{\textcolor{red}{28.3701}} & 1.3453                            & \multicolumn{1}{l|}{\textcolor{red}{0.8507}} & 0.0376                            \\ \cline{7-11} 
\multicolumn{1}{|c|}{}                                    & \multicolumn{1}{c|}{}                                         & \multicolumn{1}{c|}{}                                          & \multicolumn{1}{c|}{}                                     & \multicolumn{1}{c|}{}                                                                                                             & \multicolumn{1}{c|}{}                                                                                          & VGG                                                          & \multicolumn{1}{l|}{\textcolor{blue}{27.0698}}          & 1.3894                           & \multicolumn{1}{l|}{\textcolor{blue}{0.8464}}          & \multicolumn{1}{r|}{0.0299}       \\ \hline
\end{tabular}
\end{adjustbox}
\end{table*}
\begin{figure*}[!h]
    \centering
    \centerline{\includegraphics[width = \textwidth]{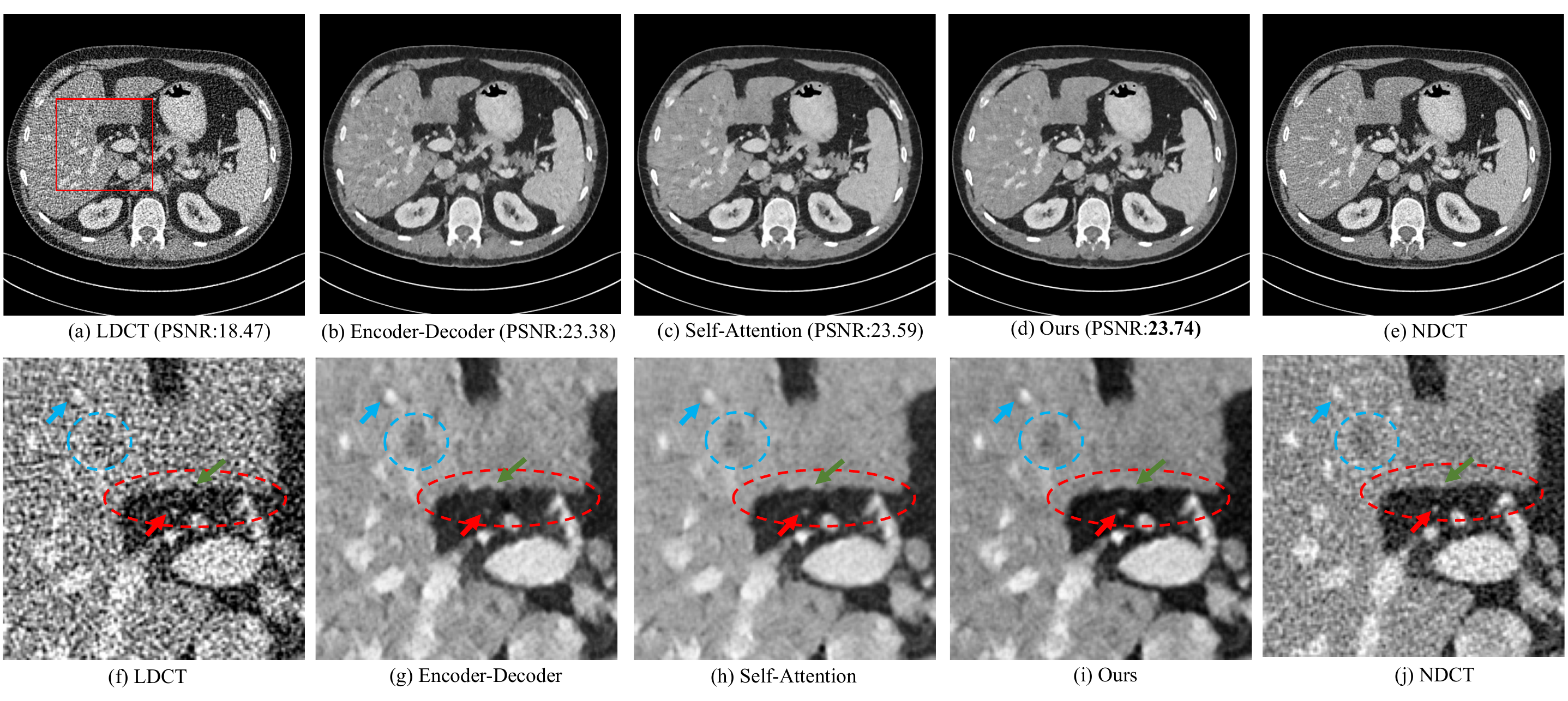}}
    \caption{The visualized comparison of overall image (in L291 case) and zoomed-in region of interest. In the zoomed-in sub-figure, the blue and red 
circles are two selected regions for better comparison.  Please zoom in for better view. The display window is [-160, 240] HU. }
    
\end{figure*}

\subsection{Ablation Study of Proposed Methods}
In this section, we perform the ablation study to validate the effectiveness of each component of the proposed GCN-MIF model, including local, non-local and contextual information.  Specifically, 3 baseline methods are constructed as following
\begin{itemize}
    \item \textbf{Baseline1}: Based on our proposed GCN-MIF model (as shown in Figure 2), the plane graph convolutional network in each GCN-MIF is removed. The modules for local and contextual information are replaced by a 3D CNN operation. Thus, Baseline1 can be regarded as a plain 3D CNN-based model, which can implicitly extract contextual information. 
    \item \textbf{Baseline2}: Based on our propsoed GCN-MIF model,
    the plane graph convolutional network in each GCN-MIF is also removed. Baseline2 aims to explore the effectiveness of depth 
    convolutional network for contextual information.
    \item \textbf{Baseline3}: Baseline3 is equal with the proposed GCN-MIF model.
\end{itemize}

In order to eliminate the impact of the loss function, each baseline methods are trained under two frameworks, including mean square error (MSE)-guided manner and WGAN-VGG one (the baseline model plays the role of generator and the structure of the discriminator follows previous work in \cite{wolterink2017generative}). We evaluate the mean and variance of the PSNR and SSIM values on the test set. 

From Table I, some observations can be noticed as below
\begin{itemize}
    \item By comparing the performance between Baseline1 and Baseline2, we can find that the explicit introduction of contextual information is indeed effective to improve the performance slightly. For example, the PSNR value increases from 28.1532 to 28.2802 under the MSE-guided framework.
    \item By analyzing the results between Baseline2 and Baseline3, we can observe that the non-local information looks more crucial for the improvement of denoising performance, as the PSNR and SSIM values achieve the best performance under two frameworks. This phenomenon may be reasonable as intra-slice information have closer  relationships with the pixel that needs to be restored. Instead, 
    the contextual information rely heavily on the thickness of the slice, which is not stable (as discussed in the section II-B).
\end{itemize}

\subsection{The comparison of different non-local models}
In this section, we explore the performance of different non-local models to show the advantages of our proposed 3D GCN module. As mentioned in Introduction section, CNN-based encoder-decoder model can extract non-local information by stacking downsampling operations. We thus construct a  CNN-based encoder-decoder model (as the \textbf{Baseline4}), which has 3 downsampling layers and corresponding upsampling layers. Similarly, the self-attention-based 
model can leverage non-local relationships. Following \cite{li2020sacnn}, the self-attention CNN (SACNN) model with 3 self-attention layers (only plane attention module) is used for comparison, as the \textbf{Baseline5}. Based on our proposed GCN-MIF model, our proposed GCN-based model can be constructed by removing the module of contextual information (depth convolutional network). In addition, each baseline method is trained under two kinds of objectives. We evaluate the computational consumption (e.g., parameters, inference time, and GPU consumption), quantitative results, and visualized comparison on test set.

As reported in Table II, one can have some observations as following
\begin{itemize}
    \item Overall, as explicit non-local models, self-attention and GCN-based models outperform encoder-decoder-based model with an obvious margin (in terms of PSNR and SSIM), which validates the effectiveness of non-local modules. 
    \item Compared with self-attention-based model, our proposed GCN-based model achieves better performance under two loss functions, which is reasonable as the proposed GCN model can leverage appropriate non-local neighbors instead of all neighbors (e.g., self-attention). The negative impact of too many non-local information may be the reason why the performance slightly decreases.
    \item For computational consumption, we can notice that our proposed GCN-based model can only take half of the running time and computational cost (e.g., GPU consumption) compared with self-attention-based model. This is obviously caused by the usage of the relationships of all the pixels for the self-attention module. Instead, the GCN module utilizes appropriate non-local information to achieve competitive performance.
\end{itemize}

The visualized comparison of denoising results (MSE-guided framework) is illustrated in the Figure 5. From the selected zoomed-in region of interest, we can notice some observations. First, for the noise suppression of the dark background area (as shown in red circle), the self-attention-based and our proposed GCN-based models have better capacity compared with encoder-decoder-based model, which may result from the introduction of non-local information as pointed by \cite{li2020sacnn}. Second, compared with self-attention-based model, our proposed GCN-based model can retain more structural information (as pointed by blue arrow), we guess that it may be caused by too many non-local information, leading to the loss of their own information. The low attenuation lesion (as shown in blue circle) is maintained by our proposed model as much as possible. Finally, our proposed method produces the sharpest edge (as pointed by green arrow), which is reasonable as the selection of non-local neighbors can be along with the edge (as shown in the first sub-figure of Figure 4). Overall, our proposed GCN-based model has a better balance between noise reduction and structure preservation.
\begin{table}[]
\renewcommand{\arraystretch}{1.2}
\centering
\caption{The quantitative comparison for different methods on the test set. For the PSNR, the
SSIM, and the VIF, the higher the better. In each training framework, the best score among all models is bolded.}
\begin{adjustbox}{width=0.9\columnwidth}
\begin{tabular}{cccc}
\hline
\multirow{2}{*}{Method} & \multicolumn{3}{c}{Evaluation Metrics}                \\ \cline{2-4} 
                        & PSNR             & SSIM            & VIF             \\ \hline
LDCT                    & 22.1708          & 0.7512          & 0.3119          \\ \hline
RED-CNN                 & \textbf{27.0296} & \textbf{0.8037} & 0.3801          \\ Encoder-Decoder-MSE                 & 26.8230 & 0.8001 & 0.3798          \\
SACNN-MSE               & 26.8909          & 0.8004          & 0.3839          \\
\textbf{Ours-MSE}                & 26.8953          & 0.8015          & \textbf{0.3860} \\ \hline
WGAN-VGG                & 25.0128          & 0.7901          & 0.3390          \\
CPCE-3D                 & 24.6051          & 0.7869          & 0.3193          \\
WGAN-SACNN-VGG               & 25.2352          & 0.7902          & 0.3392          \\
\textbf{WGAN-Ours-VGG}                    & \textbf{25.6860} & \textbf{0.8005} & \textbf{0.3449} \\ \hline
\end{tabular}
\end{adjustbox}
\end{table}
\begin{figure*}[!h]
    \centering
    \centerline{\includegraphics[width = \textwidth]{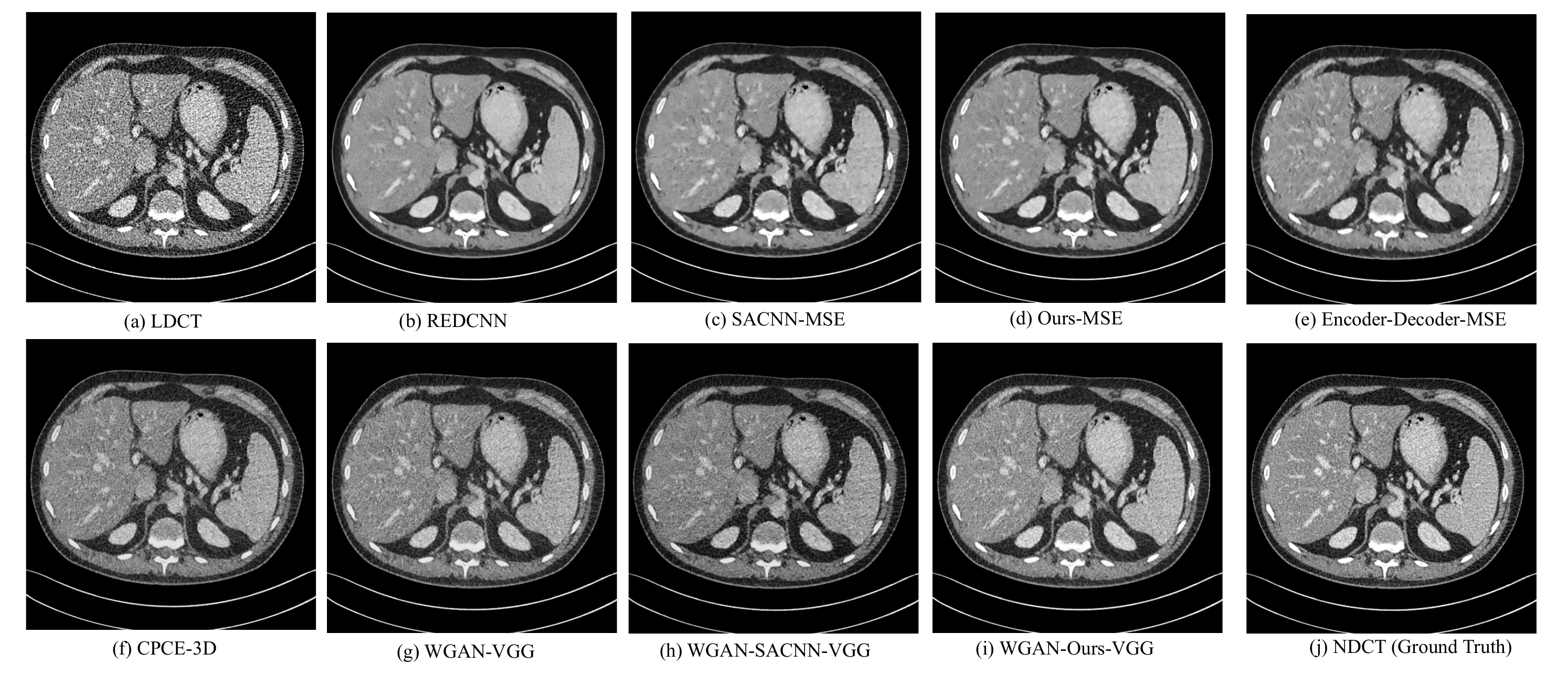}}
    \caption{The visualized comparison of overall image on test set. The display window is [-160, 240] HU.}
    
\end{figure*}
\begin{figure*}[!h]
    \centering
    \centerline{\includegraphics[width = \textwidth]{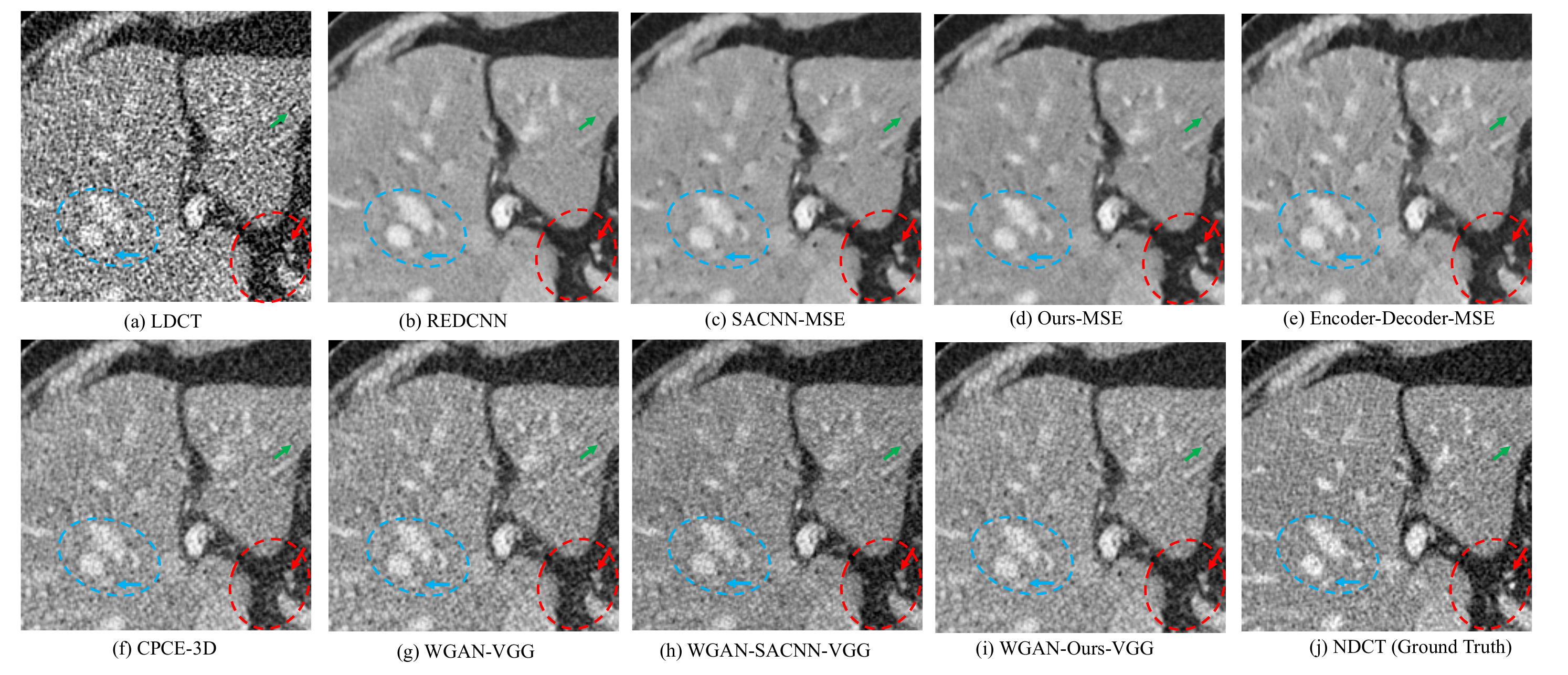}}
    \caption{The visualized comparison of zoomed-in region of interest. In the zoomed-in sub-figure, the blue and red circles
are two selected regions for better comparison. Please zoom in for better view. The display window is [-160, 240] HU.}
    
\end{figure*}
\subsection{The comparison of different LDCT denoising models}
In this section, we compare our proposed GCN-MIF model with baseline methods on test set. For quantitative performance of denoising results, PSNR, SSIM and visual information fidelity (VIF) \cite{sheikh2005visual} are used.The baseline methods include RED-CNN, encoder-decoder-based model (i.e., Baseline4), SACNN, WGAN-VGG \cite{wolterink2017generative}, and CPCE-3D models. Specially, our proposed GCN-MIF model and SACNN model are trained by two kinds of objectives (including MSE-guided manner and WGAN-VGG framework). 

As reported in Table III, under the guidance of MSE, RED-CNN achieves the highest PSNR and SSIM among all methods, which is reasonable as RED-CNN usually generates over-smooth denoising results. It is worthy noting that our proposed method not only has the second highest PSNR and SSIM, but also achieves the best VIF (which can reflect human visual system by using natural statistics models). Thus, the proposed GCN-MIF model balances the noise reduction and structure preservation well.
For WGAN-VGG-based models, our proposed model outperforms others with an obvious margin.

The visualized comparison of denoising results is illustrated in the Figure 6.
We can roughly observe from Figure 6 that the denoising results generated by our proposed GCN-MIF model (including Our-MSE and WGAN-Ours-VGG) are brighter than others, which reflects the preserving of structural information and contrast ratio. 

The zoomed-in region of interest is also presented in the Figure 7.  One can have some observations. For MSE-guided models, our proposed model retains more structural information compared with RED-CNN and SACNN-MSE models. For example, the blood vessel (as pointed by green arrow) and subtle detail (as pointed by blue arrow) can be still noticed by our proposed method. Although encoder-decoder-MSE model achieves good structure preserving,
the ability of noise suppression is very limited (as shown in red circle). For WGAN-VGG-based models, due to the introduction of explicit non-local information, SACNN and our proposed model have better performance for noise reduction (as circled in the dark background area). Compared with SACNN, our proposed model maintains the subtle structure more. As pointed by blue circle, the denoising result by SACNN is difficult to be noticed.

\section{Double-blind Scoring Experiment}

\subsection{The Test Set for Double-Blind Scoring Experiments.}
To adequately compare the denoising performance of our proposed model with that of other representative models, a paired  low/normal-dose clinical CT dataset with various examination regions, i.e., Low-Dose CT images and Projection Data (a.k.a., LDCT-PD) dataset \footnote{https://wik
i.cancerimagingarchive.net/pages/viewpage.action?\\pageId=52758026}, is used in this study. The LDCT-PD dataset was released by Mayo Clinic, including 99 non-contrast head  CT scans acquired for the patients of acute cognitive or motor deficit, 100 low-dose non-contrast chest scans acquired to screen high-risk patients for pulmonary nodules, and 100 contrast-enhanced CT scans of the abdomen acquired to look for the patients of metastatic liver lesions.  Similar to  \cite{RN314}, the chest and abdomen CT images are selected for double-blind study. Every part consists of 67 patients' scans. The normal-dose chest and abdomen CT images are scanned under the settings of 120kV, 250mAs and 100kV, 300mAs, repectively. The corresponding low-dose images are 10$\%$  and 25$\%$ of normal-dose. It should be noted that the LDCT-PD dataset provided a well-written clinical report, including the labeled locations of the lesion and the diagnosis types of the lesion. Based on this, we thus select a CT slice labeled with lesion and its adjacent slices (a lesion slice and its front 4 slices and rear 4 slices, totally 9 CT slices) to denote overall slices of a patient. We then can study the influence of different denoising results for the lesion.

There are two points that need to be noted. First, the imaging parameters, equipment vendors and acquired locations have  differences between the training set and test set, which well reflects the complex environments in clinic. To obtain the competitive denoising performance, the deep learning-based models must have a good adaptive and generalized capacity. This is one of the motivations for the double-blind test that if the deep learning-based denoising models have the effectiveness in clinic, and which models have the best denoising ability under  complex conditions. Second, the test images in our double-blind experiment refer to plenty of CT images with labeled lesions, which is very meaningful for the  radiologists, allowing them to evaluate if the denoising results will influence the judgment of lesion type or image feature. 

\subsection{The details of double-blind reader study}
The low-dose CT images of each selected patient are denoised by three deep learning-based models (MAP-NN, RED-CNN, and our proposed GCN-MIF). Note that each denoising result generated by MAP-NN model has 5 sub-figures beforehand, which corresponds to 5 (1 to 5) depth parameters, in order to simplify operations. By this way, radiologists do not need extra waiting time for model inference. By communicating with them, radiologists prefer the denoising results with 3 stacked modules, as the corresponding images can typically balance the noise suppression and structure preserving well. 

For each patient, 5 sub-folders can be obtained, i.e., 3 sub-folders with different denoising results, LDCT sub-folder, and NDCT sub-folder. The sub-folders of denoising results are named randomly (such as measure1, measure2, and measure3). Finally, the total 134 folders are obtained for double-blind study. Three experienced radiologists (J. Shen, radiologist $\#1$ with 23 years experience, D. Wu, radiologist $\#2$ with 10 years experience, X. Pan, radiologist $\#3$ with 14 years experience) participate in this experiment. For the standard of evaluation, we adopt the scheme of  4-point scale as in the previous study \cite{kim2007antibiofouling} in terms of image noise, structural fidelity and overall score. Specifically, 1 score denotes the quality of the CT image as unacceptable for clinical diagnosis. 2 score denotes the CT image can  provide limited diagnostic information only. 3 score denotes that the CT image is acceptable and can provide the average diagnostic information. 4 score denotes the CT image has a good quality in terms of the accurate diagnosis and interpretation. Unlike any other systemic double-blind study in \cite{kim2007antibiofouling}, we add the overall score as a part of the evaluations. As some results may have very good denoising performance in terms of  the ROIs but be sub-optimal in terms of the non-ROIs, the radiologists thus can reflect the real image quality using the term of overall score.

\subsection{Results}

% Please add the following required packages to your document preamble:
% \usepackage{multirow}
% Please add the following required packages to your document preamble:
% \usepackage{multirow}

% Please add the following required packages to your document preamble:
% \usepackage{multirow}

% Please add the following required packages to your document preamble:
% \usepackage{multirow}

To compare the  performance of representative deep learning-based methods, a double-blind study is carried out in this paper. The representative methods for comparison include RED-CNN \cite{chen2017Low2}, MAP-NN \cite{RN314}, and our proposed GCN-MIF model. The corresponding details can be found in Table IV. Using mean square error (MSE) as the loss function, RED-CNN is widely adopted  due to its well noise suppression. Through systemic double-blind study, MAP-NN shows competitive denoising performance compared with commercial iterative reconstruction methods.  To be fair, all methods are trained on AAPM-Moyo dataset, which includes 10 patients' abdomen scans (paired normal-dose and low-dose (25$\%$ of normal-dose) CT images). We utilize a separate  dataset, i.e., LDCT-PD dataset, to evaluate the denoising capacity of different methods in real-world complex environments. In LDCT-PD dataset, the selected test data totally includes 67 pairs of normal/low-dose (25$\%$ of normal-dose) abdomen scans and 67 pairs of normal/low-does (10$\%$ of normal-dose) chest scans. It can be easily noted that the dose level of low-dose chest scans in test set is extremely lower than that of the training one. These test models thus must have an adaptive and generalized capacity to handle this various environment, which we will analyze further later in this paper. 

Unlike previous double-blind studies, the denoising performance of different methods is completely evaluated on the region of lesion. We believe that this may be more valuable for clinical environments. Specifically, the abdomen and chest CT images have the lesion of metastatic liver and the lesion of pulmonary nodule, respectively. Three experienced radiologists from West China Hospital and West China No. 4 Hospital  participated in this study. They are named as "radiologist $\#1$", "radiologist $\#2$", and "radiologist $\#3$", respectively. The standard of 4-point scale is used to evaluate the denoising performance in terms of noise, fidelity, and overall score. Briefly, the higher the score, the better. More details about the settings of double-blind study can be found in the section  \hyperref[GCN]{Method}.
\begin{figure*}
  \centering
  
  \subfigure{
  \begin{minipage}{0.48\linewidth}
  \centering

     \includegraphics[width=\linewidth,height=0.66\linewidth]{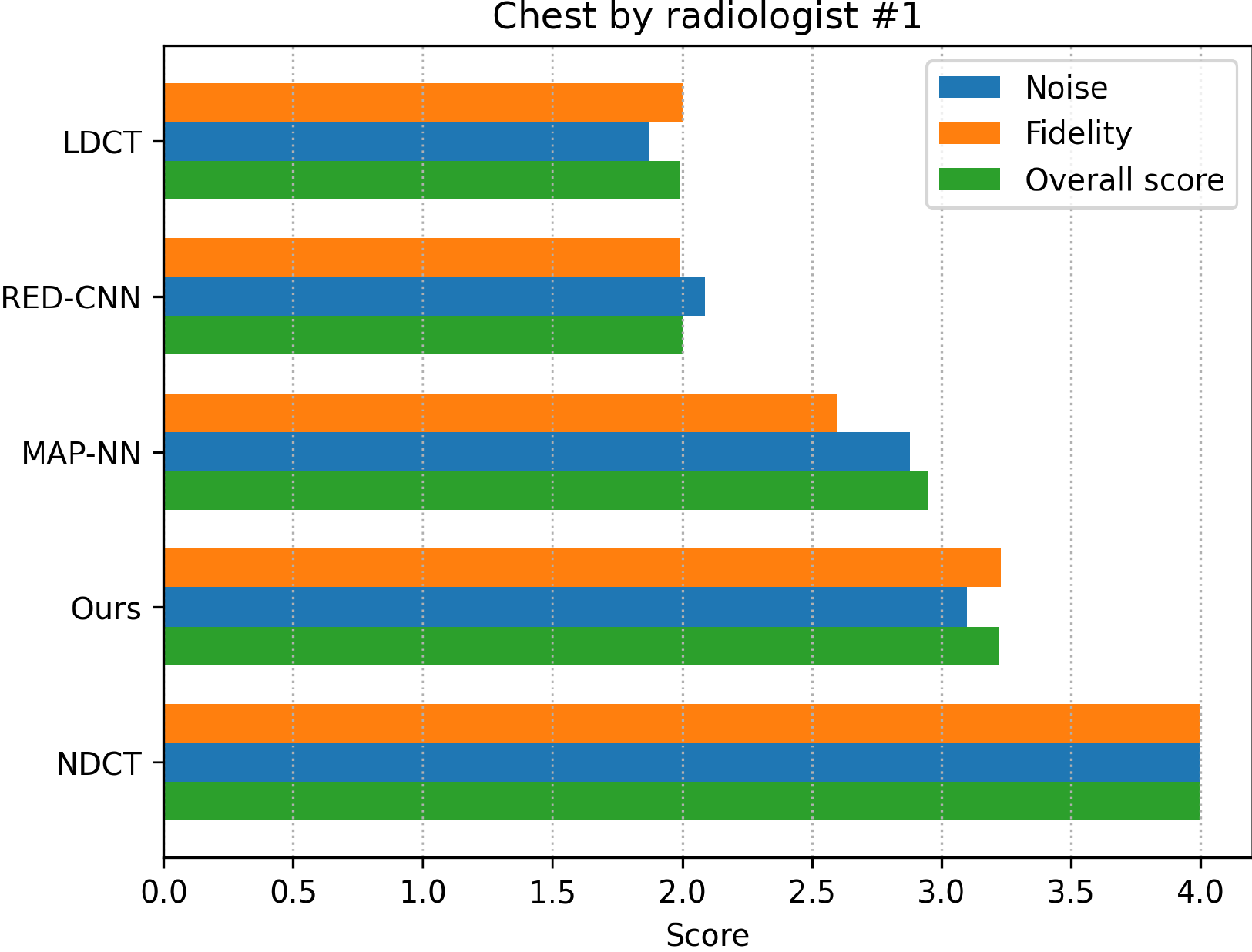}%{lilun1.eps}
  \end{minipage}
  }
  \subfigure{
  \begin{minipage}{0.48\linewidth}
  \centering

     \includegraphics[width=\linewidth,height=0.66\linewidth]{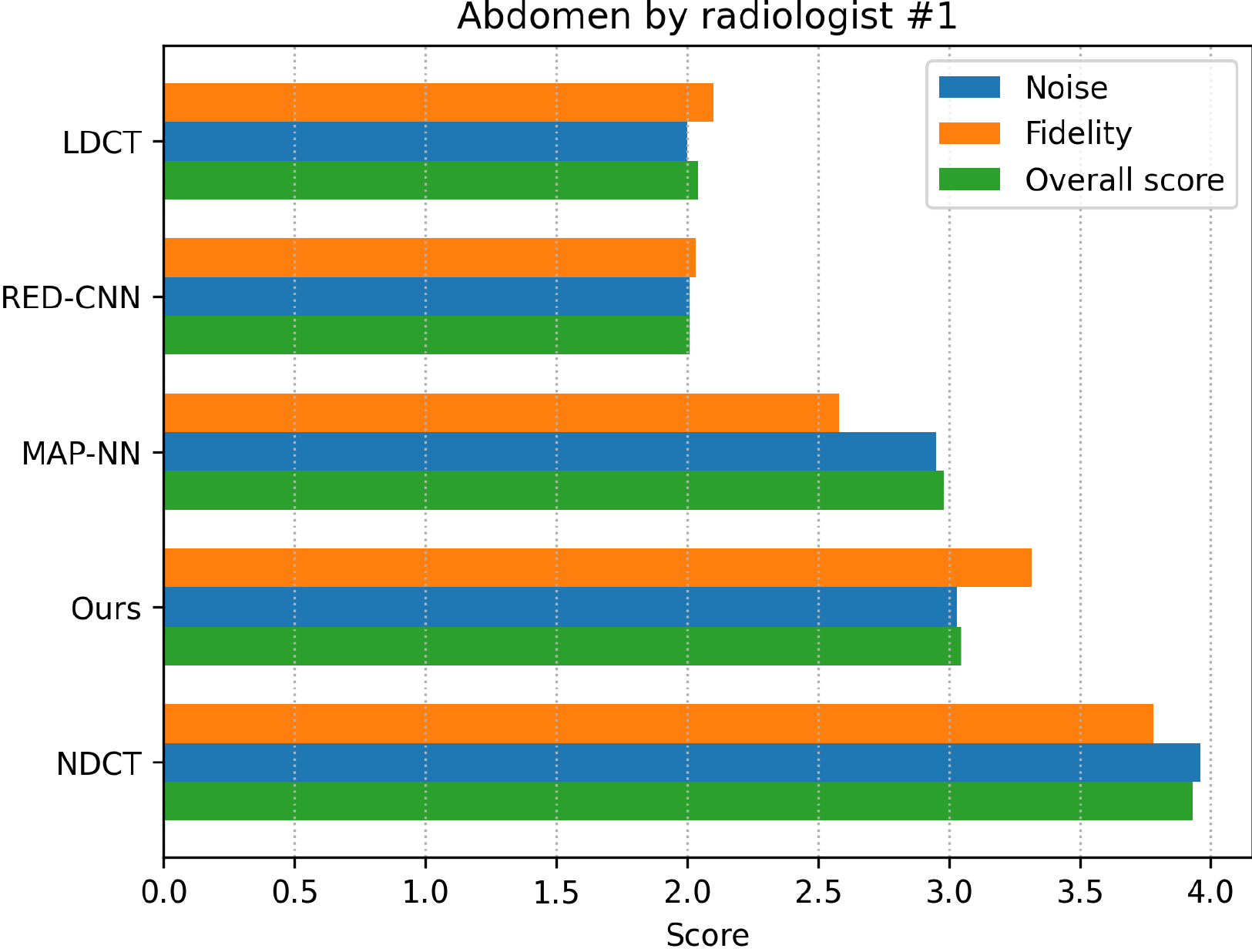}%{shiji1.eps}
  \end{minipage}
  }
   
  \subfigure{
  \begin{minipage}{0.48\linewidth}
  \centering

     \includegraphics[width=\linewidth,height=0.66\linewidth]{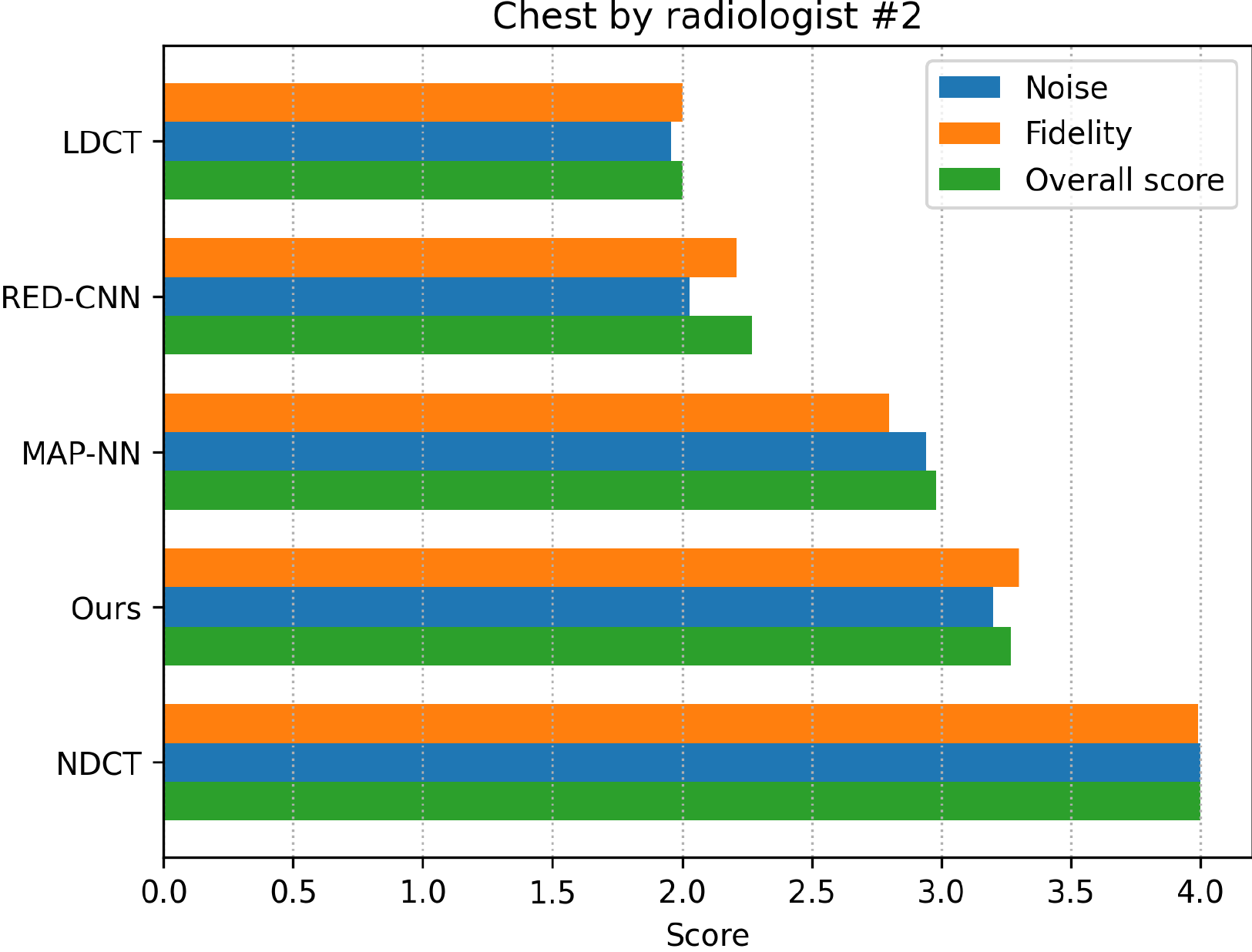}%{lilun1.eps}
  \end{minipage}
  }
  \subfigure{
  \begin{minipage}{0.48\linewidth}
  \centering

     \includegraphics[width=\linewidth,height=0.66\linewidth]{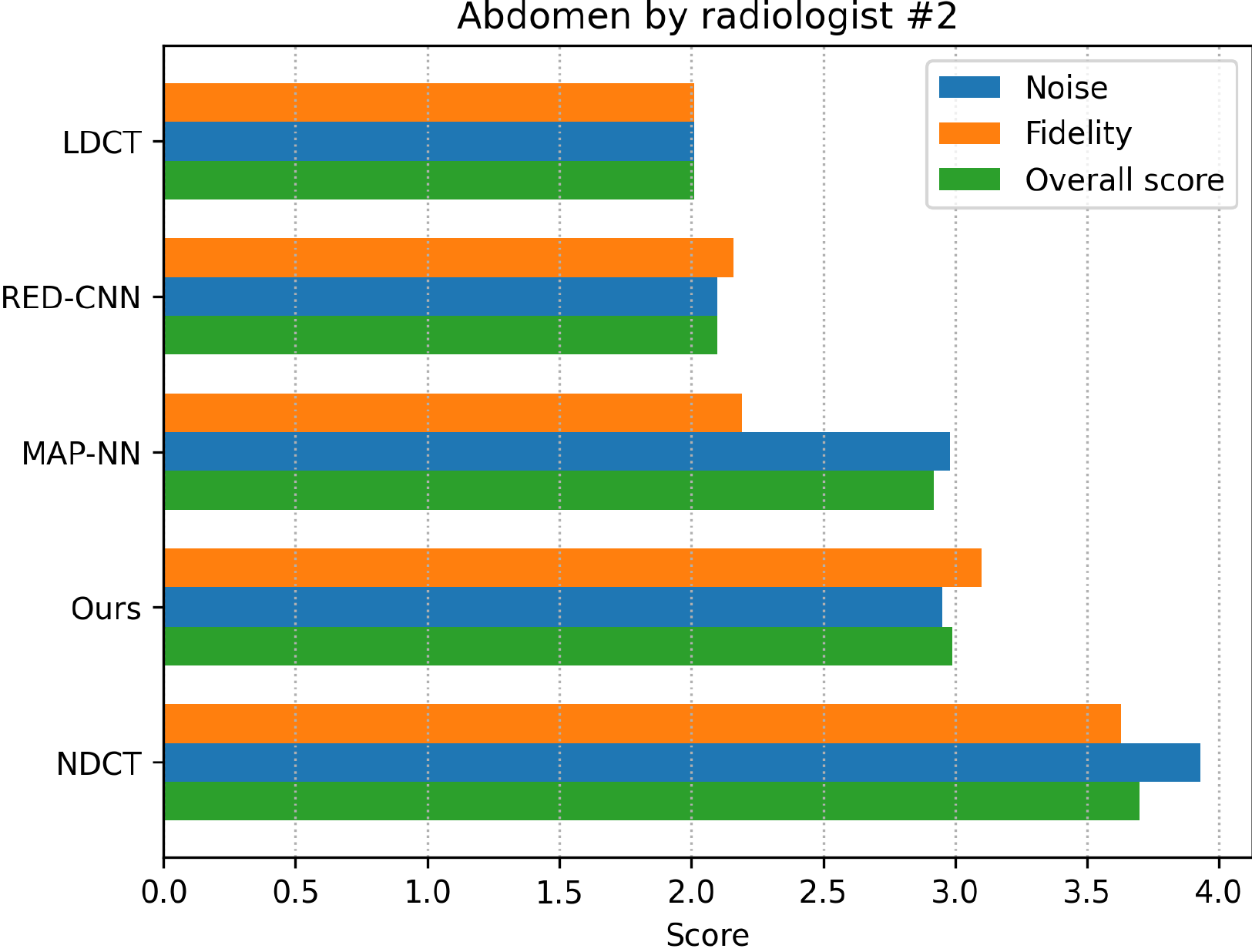}%{shiji1.eps}
  \end{minipage}
  }
  
  \subfigure{
  \begin{minipage}{0.48\linewidth}
  \centering

     \includegraphics[width=\linewidth,height=0.66\linewidth]{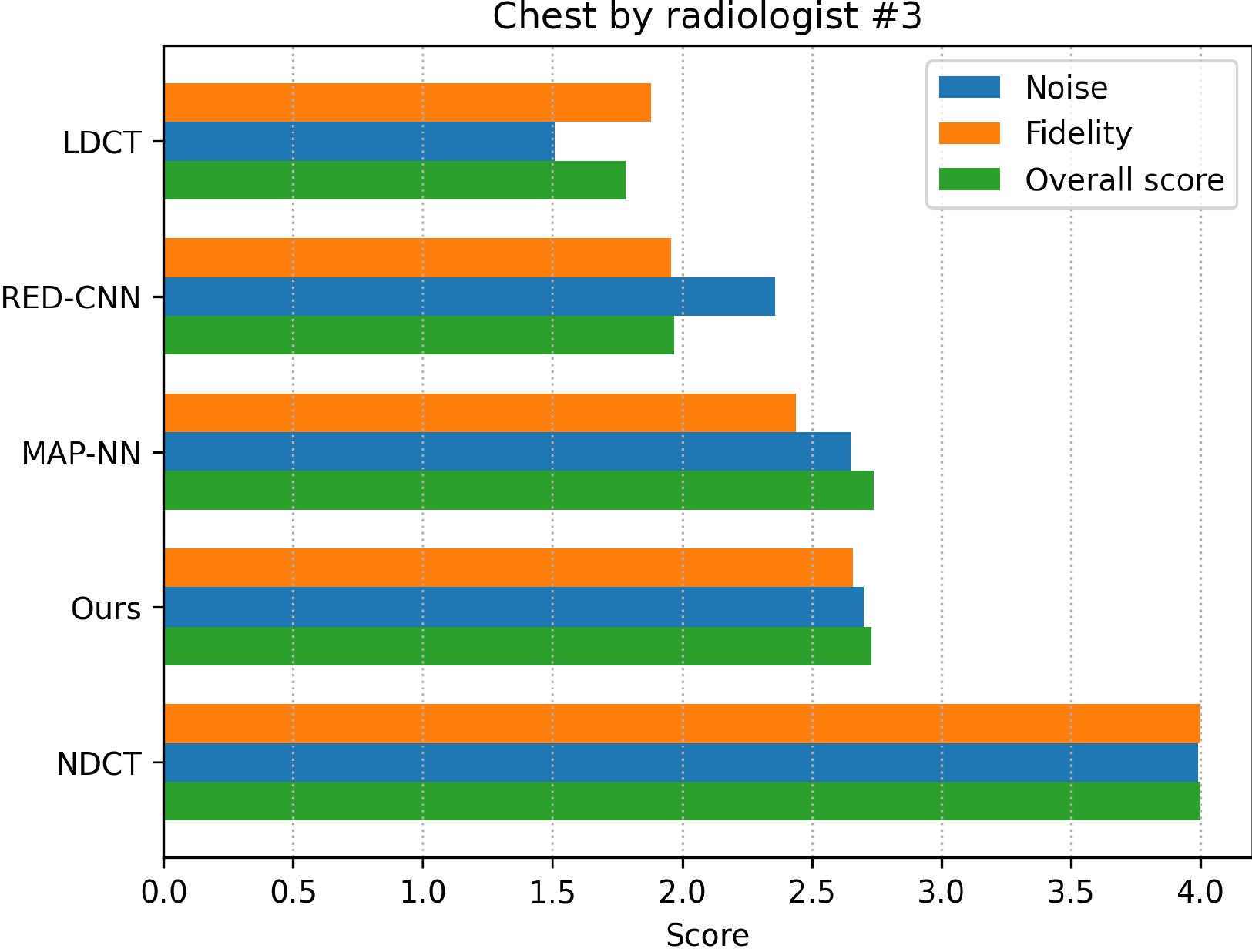}%{lilun1.eps}
  \end{minipage}
  }
  \subfigure{
  \begin{minipage}{0.48\linewidth}
  \centering

     \includegraphics[width=\linewidth,height=0.66\linewidth]{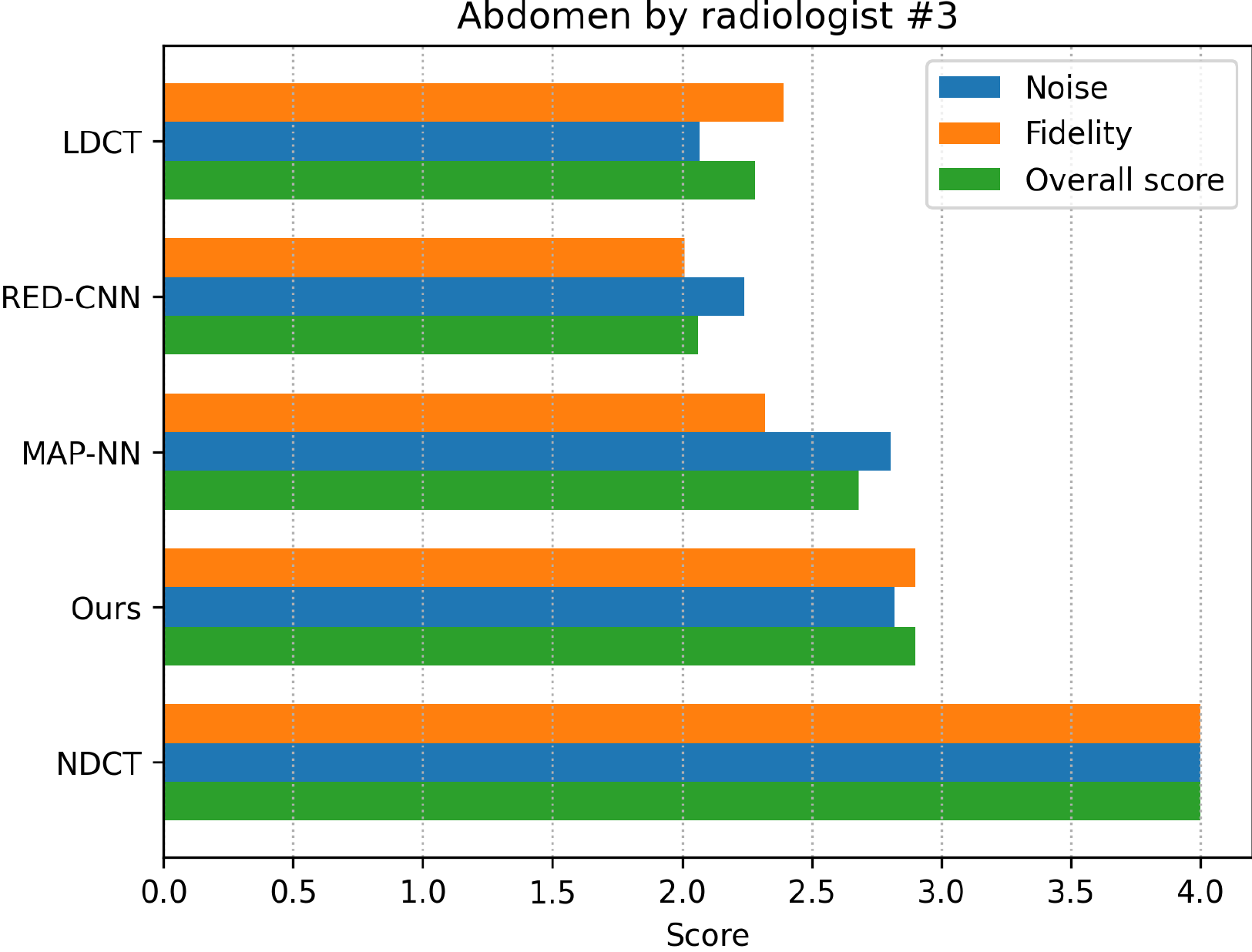}%{shiji1.eps}
  \end{minipage}
  }
  \caption{The results of double-blind study by three radiologists aspect to the chest and the abdomen. For the noise, fidelity, and overall score, the higher the better. The full score is 4 for every evaluation standard.}
\label{DOUBEL_1}
\end{figure*}

\textbf{The analysis of the results for the scans of the chest.} 
As illustrated in the first column of Figure \ref{DOUBEL_1}, we can make four important observations. First, our proposed model significantly outperforms other methods in terms of noise, fidelity, and overall score, except for a slight improvement in two results (noise and fidelity) from radiologist $\#3$. This shows the superiority of integrating the graph convolutional network. Second, we can notice that RED-CNN is significantly behind MAP-NN and our proposed model. Although the framework of the MSE-guided (adopted by RED-CNN) would intuitively generate very smooth and little noisy results (see the Figure \ref{visualized_1_blind} and Figure \ref{visualized_2_blind}), the radiologist may not favour
this style that is much limited for the improvement of the  diagnosis. Instead,  the results of generative adversarial network (GAN) framework (adopted by MAP-NN and ours) make an obvious improvement compared with LDCT. Third, compared with MAP-NN, our proposed model still achieves better performance despite using the same framework. Fourth, the better results reflect that our proposed model has better adaptive and generalized capacity, because the dose level of the test set mismatches that of the training set, being very much lower. This property will take a strong potential in complex clinical environments, especially for various  imaging parameters, equipment vendors and acquired locations.  

\begin{table}
\centering
\renewcommand{\arraystretch}{1.4}
\caption{The details of deep learning-based methods for comparison}
\begin{adjustbox}{max width=\columnwidth}
\begin{tabular}{cccc}
\hline
Method                                                           & \begin{tabular}[c]{@{}c@{}}RED-CNN\\  {[}TMI,2017{]}\end{tabular} & \begin{tabular}[c]{@{}c@{}}MAP-NN \\ {[}Nature MI,2019{]}\end{tabular} & \begin{tabular}[c]{@{}c@{}}GCN-MIF \\ (Ours)\end{tabular} \\ \hline
Backbone                                                         & CNN                                                               & CNN                                                                    & GCN+CNN                                                \\ 
 \hline
Traning Set                                                      & AAPM-Mayo                                                         & AAPM-Mayo                                                              & AAPM-Mayo                                              \\ \hline
\end{tabular}
\end{adjustbox}
\end{table}
\textbf{The analysis of the results for the scans of the abdomen.}
The second column of Figure \ref{DOUBEL_1} shows the results of evaluation for the abdomen. One has the following observations: First, putting together the results of three metrics, our proposed model achieves the best performance, especially for the term of fidelity. This also benefits greatly from the superiority of the graph convolutional network. The proposed model leverages comprehensive information, which is  intuitively useful for the structure preservation compared with single information (such as RED-CNN and MAP-NN). Second, as the lesion of metastatic liver is usually characterized as the closely black region, the fidelity of grayscale and shape thus is very important for the diagnosis. Interestingly, the fidelity evaluation of our proposed model  significantly outperforms other models in all radiologist's results. This will contribute to the diagnosis of some unobservable lesions.  
\begin{figure*}
  \centering
  
  \subfigure{
  \begin{minipage}{0.48\linewidth}
  \centering

     \includegraphics[width=\linewidth,height=0.66\linewidth]{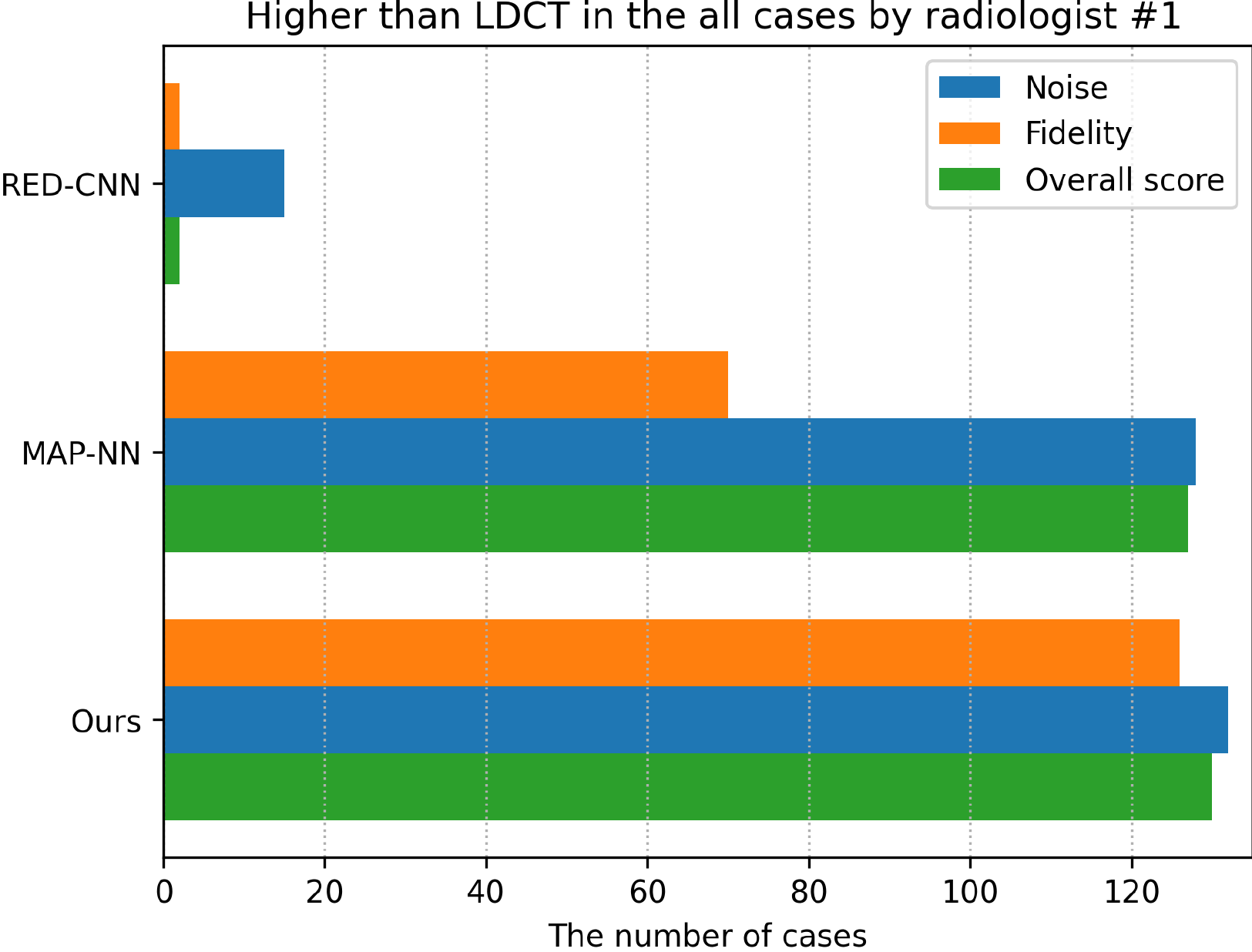}%{lilun1.eps}
  \end{minipage}
  }
  \subfigure{
  \begin{minipage}{0.48\linewidth}
  \centering

     \includegraphics[width=\linewidth,height=0.66\linewidth]{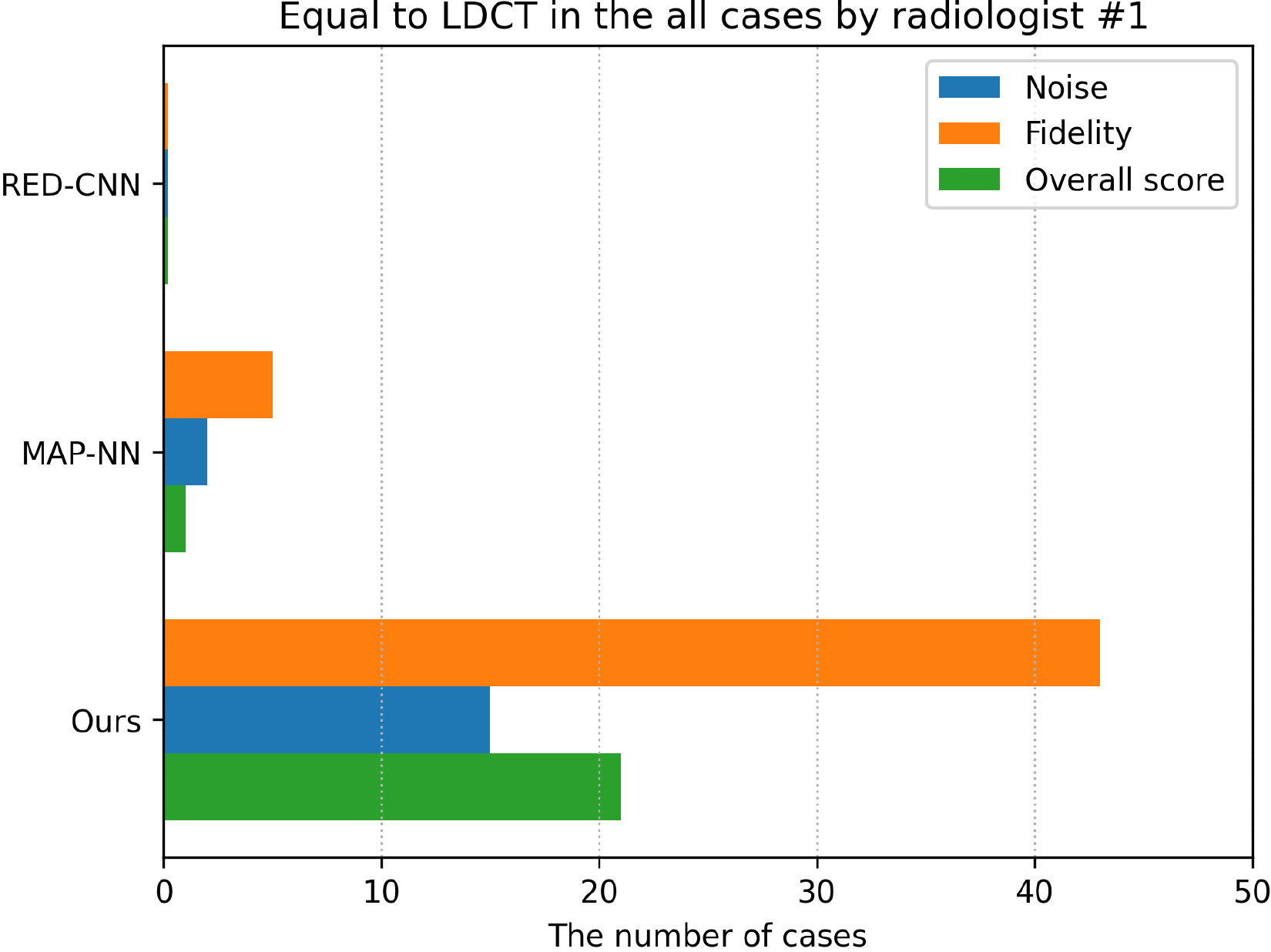}%{shiji1.eps}
  \end{minipage}
  }
   
  \subfigure{
  \begin{minipage}{0.48\linewidth}
  \centering

     \includegraphics[width=\linewidth,height=0.66\linewidth]{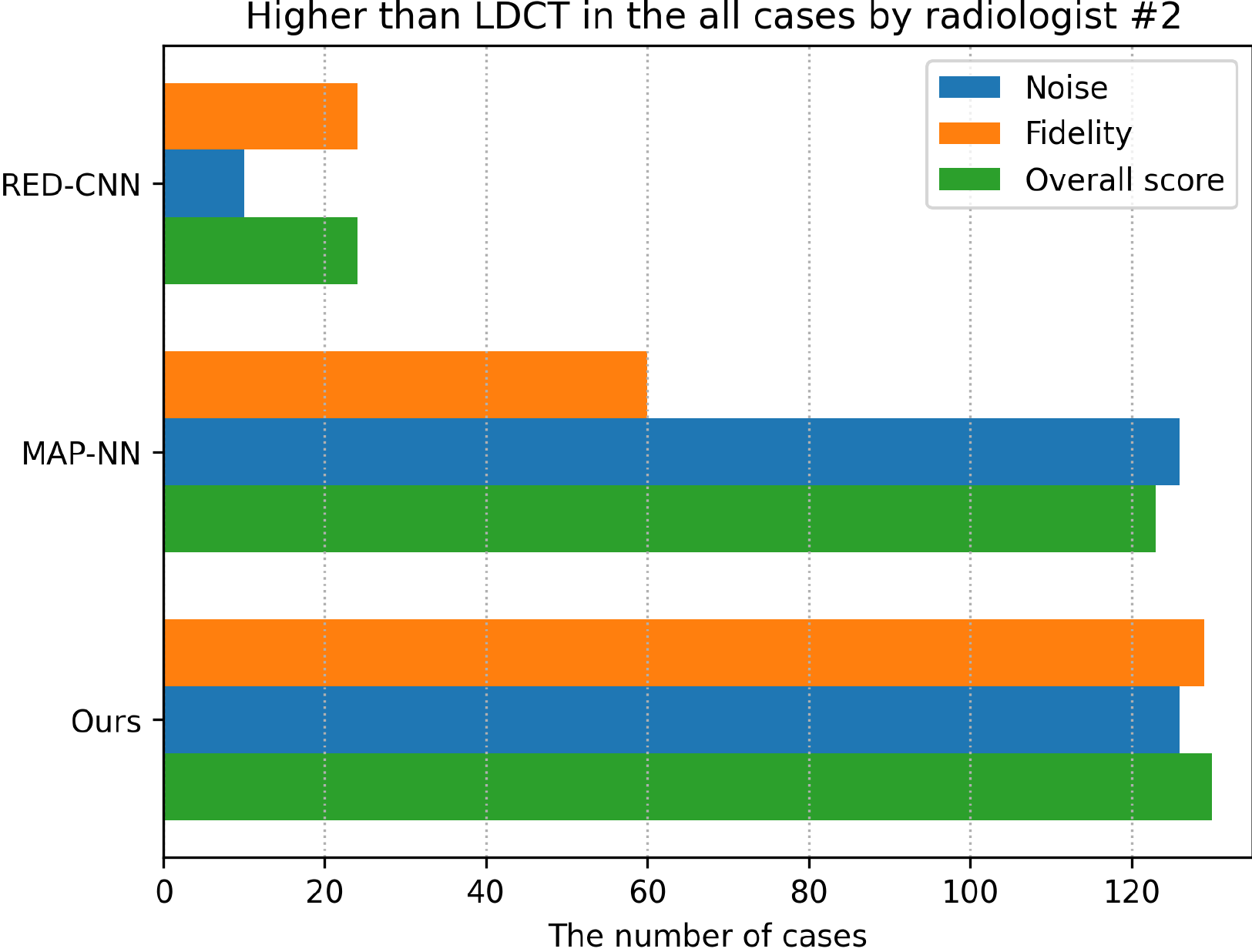}%{lilun1.eps}
  \end{minipage}
  }
  \subfigure{
  \begin{minipage}{0.48\linewidth}
  \centering

     \includegraphics[width=\linewidth,height=0.66\linewidth]{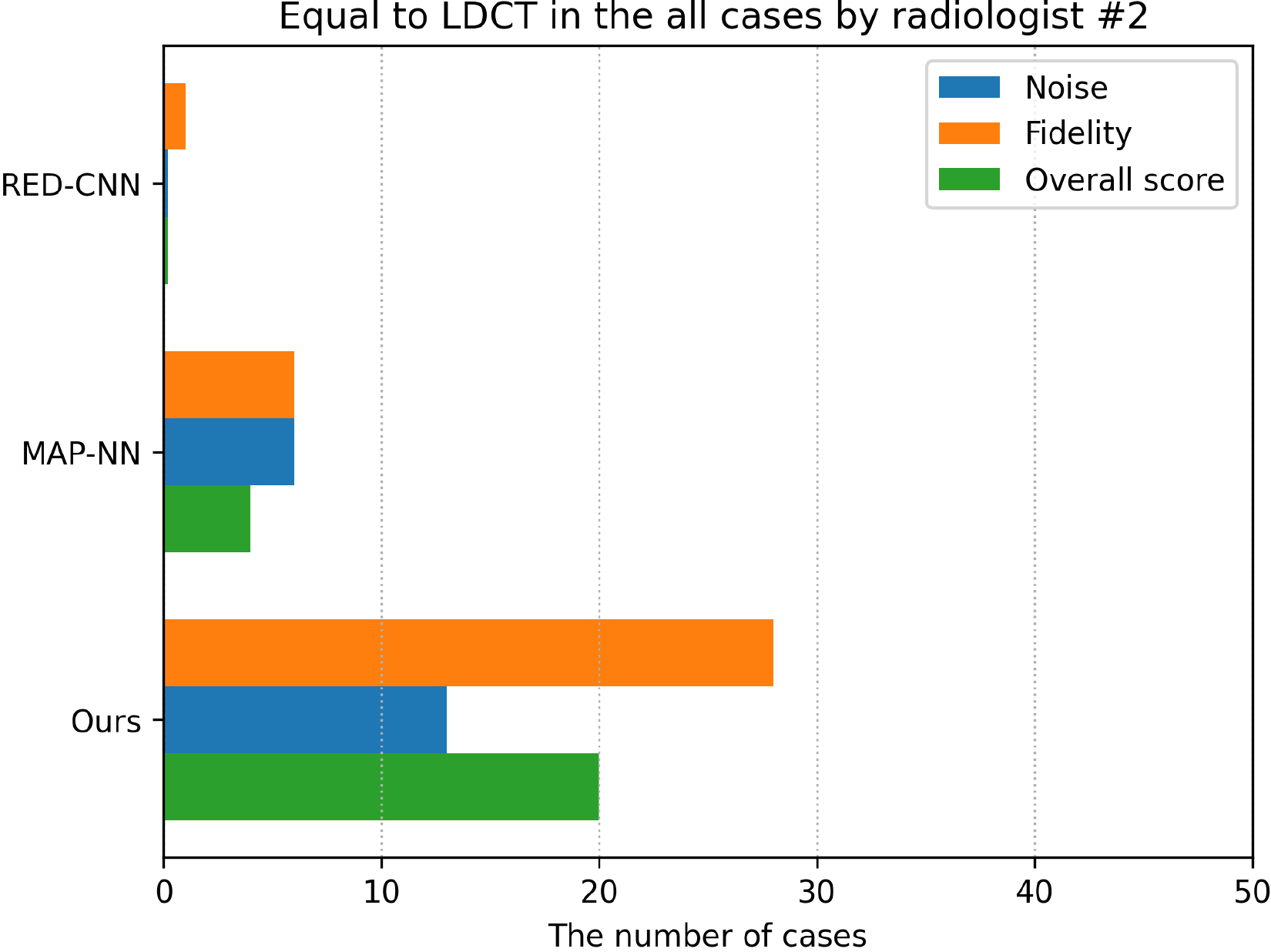}%{shiji1.eps}
  \end{minipage}
  }
  
  \subfigure{
  \begin{minipage}{0.48\linewidth}
  \centering

     \includegraphics[width=\linewidth,height=0.66\linewidth]{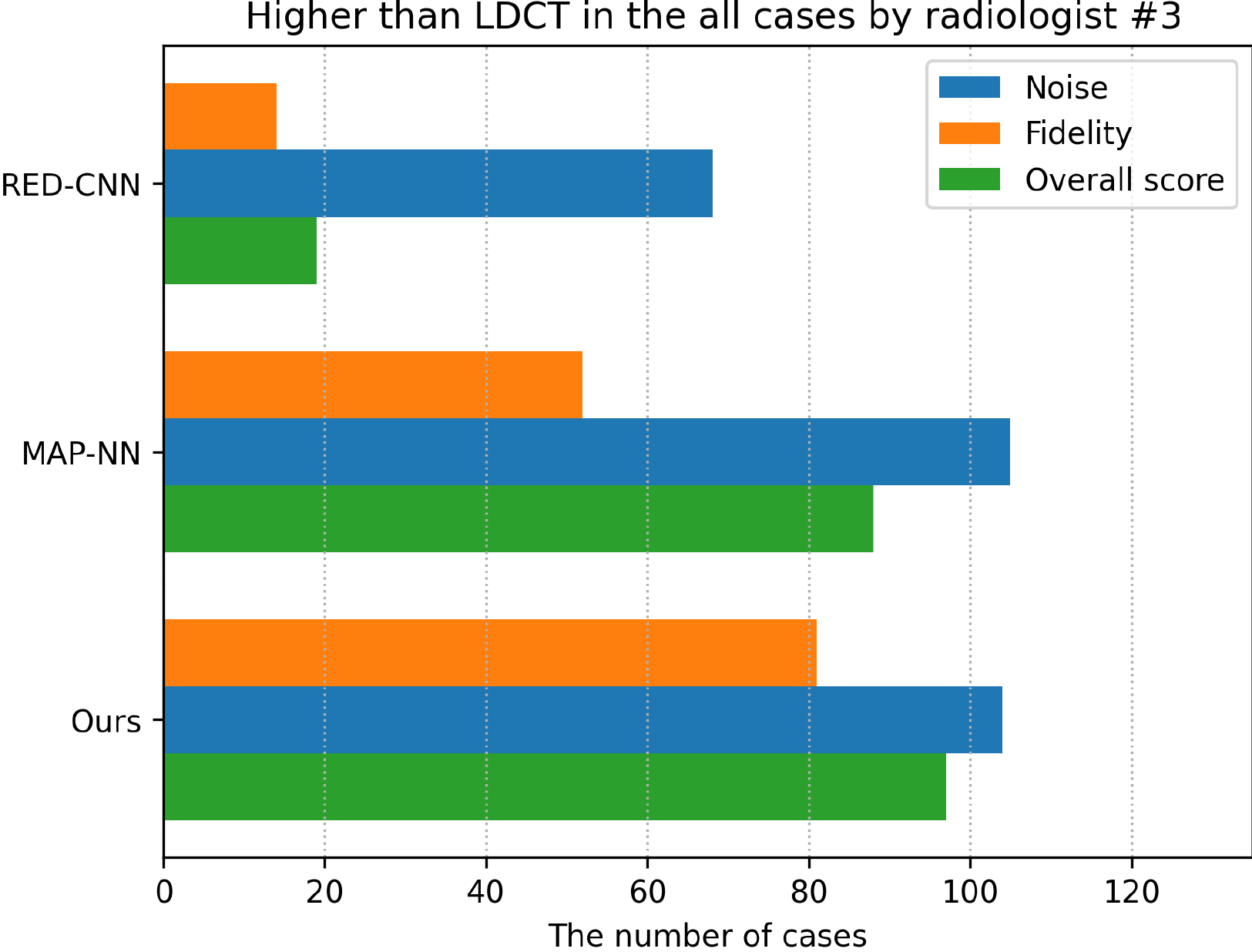}%{lilun1.eps}
  \end{minipage}
  }
  \subfigure{
  \begin{minipage}{0.48\linewidth}
  \centering

     \includegraphics[width=\linewidth,height=0.66\linewidth]{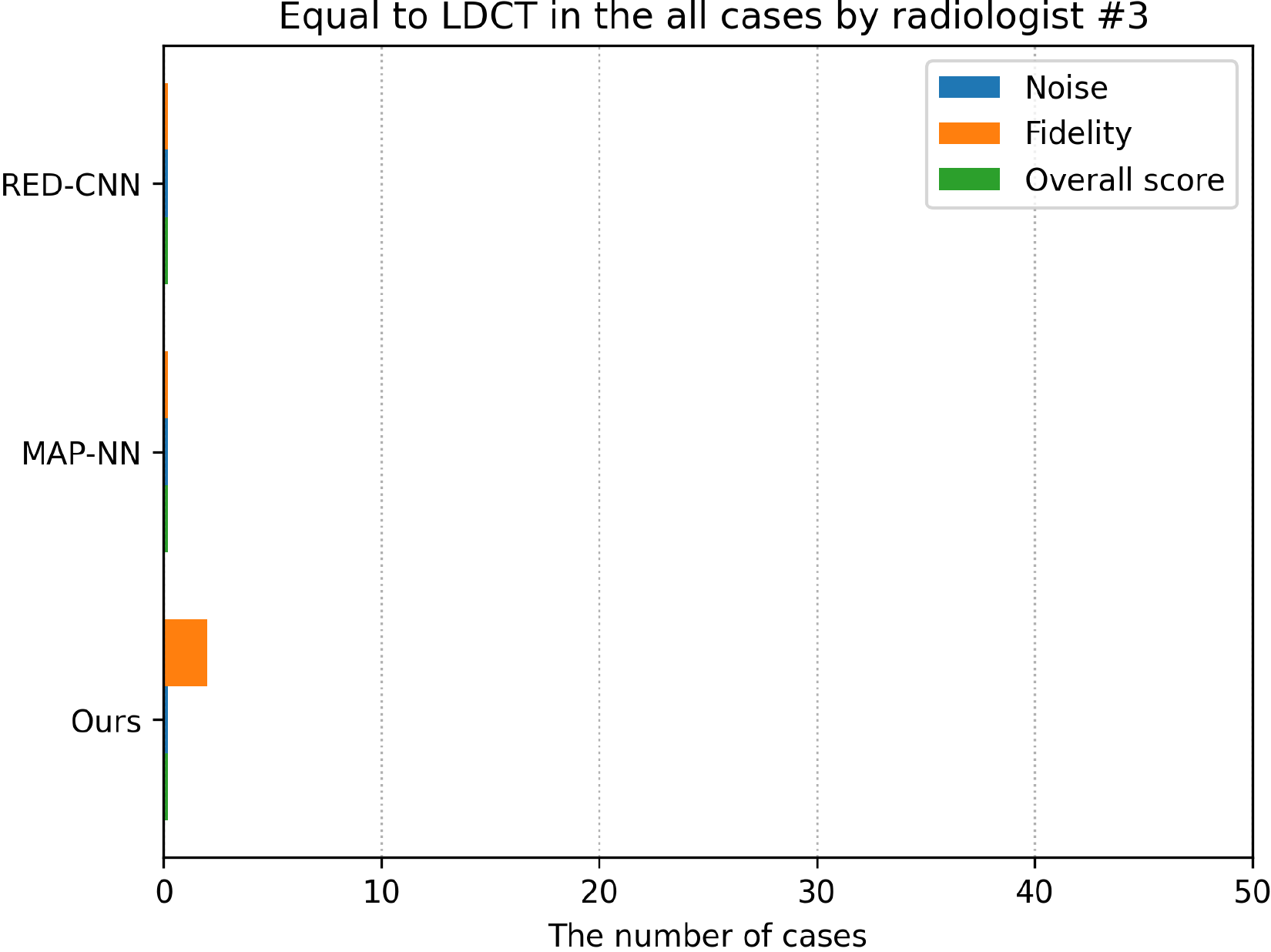}%{shiji1.eps}
  \end{minipage}
  }
  \caption{The results of double-blind study by three radiologists aspect to all cases. }
\label{double_2}
\end{figure*}

\begin{figure*}

    \centering
    \centerline{\includegraphics[width = \textwidth]{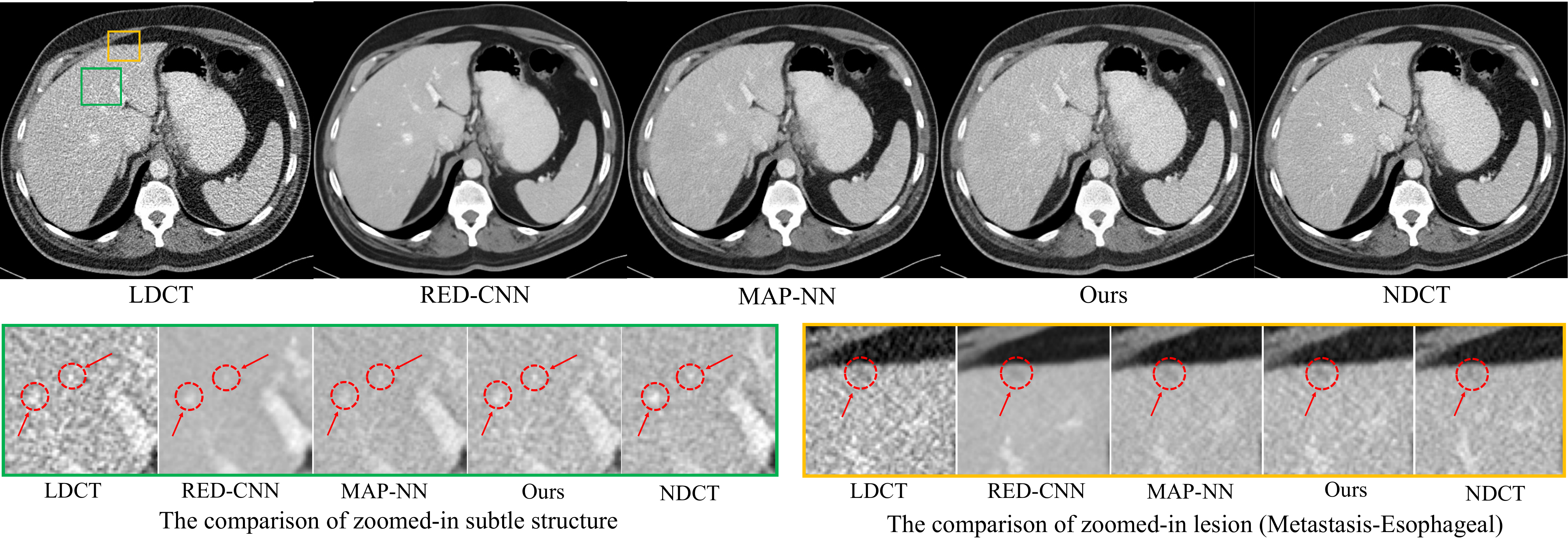}}
    \caption{The visual comparisons for zoomed-in subtle structure and lesion, in term of an example of the abdomen. The green box and yellow box are the region of zoomed-in subtle structure and the region of zoomed-in lesion, respectively. The red circles and arrows are the suggested region for comparison.  }
    
    \label{visualized_1_blind}
\end{figure*}

\begin{figure*}

    \centering
    \centerline{\includegraphics[width = \textwidth]{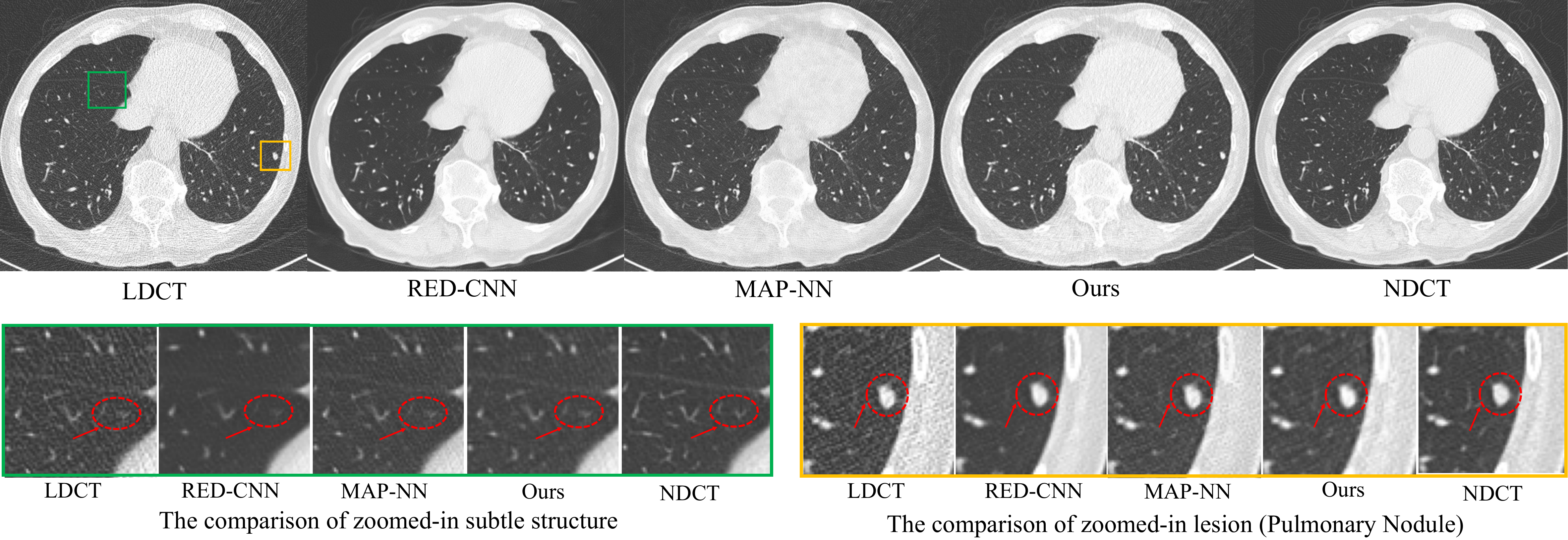}}
    \caption{The visual comparisons for zoomed-in subtle structure and lesion, in terms of an example of the chest. The green box and yellow box are the region of zoomed-in subtle structure and the region of zoomed-in lesion, respectively. The red circles and arrows are the suggested regions for comparison.}
    \label{visualized_2_blind}
\end{figure*}
\textbf{The analysis of the results higher than LDCT in all cases.} The number of cases with scores higher than LDCT images is also important to indicate the overall performance of denoising results for all cases, except the average score of different methods for different body regions. As shown in the first column of Figure \ref{double_2}, we have some observations. First, in terms of results of radiologist $\#1$ and $\#2$, we can find that our proposed model achieves improvement for all cases basically regardless of the noise, the fidelity, and the overall score.  Although MAP-NN achieves good performance for the noise and overall score (our proposed model still has a slight superiority), the improvement of fidelity can not cover all cases. Intuitively, it is very hard to balance the noise removal and structure fidelity. The results of MAP-NN clearly show this challenge.
Instead, our proposed model enjoys extra non-local  and contextual information, which shows that it is more suitable to handle this challenge. Second, due to a possible preference among radiologists, the improvements do not cover all cases in the radiologist $\#3$'s results. However, our proposed model still  significantly outperforms others in terms of the fidelity. 

\textbf{The analysis of the results equal to NDCT in all cases.} We are interested in how many denoising cases achieved the level of NDCT images, which can further represent the denoising capacity. Our proposed model achieves the best performance by analyzing all radiologist's results, as shown in the second column of Figure \ref{double_2}. Interestingly, in the radiologist $\#3$'s results, our proposed model is the only one that can achieve the level of NDCT images for the fidelity. In summary, our proposed model has the greatest potential to achieve the quality of NDCT images.

\textbf{The analysis of visual results for zoomed-in subtle structure and lesion.}
\textit{Abdomen:} As shown in Figure \ref{visualized_1_blind}, we can make some important observations. First, the green box shows the comparison of zoomed-in subtle structure, we can find  these structures (within red circles) nearly disappear for RED-CNN, due to the easily observed over-smoothness. Similarly, MAP-NN suffers from the same problem and also has a slight over-smoothness. Instead, our proposed model greatly preserves the subtle structure and generates the texture closest to NDCT. Both RED-CNN and MAP-NN only leverage the local information by the CNN, so it is difficult to balance the local detail and noise removal with limited information. Our proposed model adopts the framework of multi-information fusion such that extra information can be used as a supplement to produce the optimal results regardless of structure and texture.
Second, as illustrate in the red circle of yellow box in Figure \ref{visualized_1_blind}, our proposed model has the most obvious observation for the lesion (diagnosed as Metastasis-Esophageal), especially for the level of the grayscale. However, the lesion in RED-CNN and MAP-NN becomes very fuzzy. The superiority of lesion region further proves the effect of multi-information fusion framework. \textit{Chest:} As shown in the yellow box of Figure \ref{visualized_2_blind}, we can find that all models achieve the preservation of lesion (within the red circle). However, MAP-NN has more easily observed noise points compared with RED-CNN and our proposed model. RED-CNN  losses the subtle structure basically as shown in red circle of green box. In summary, our proposed model has the most impressive visual performance, which naturally obtains the best performance of double-blind study as reported in Figure \ref{DOUBEL_1} and \ref{double_2}.
\section{Discussion}
Compared with existing CNN-based LDCT denoising methods, the superiority of our proposed method can be summarized in four parts. First, our proposed model explicitly introduces the non-local and contextual information, which can contribute better denoising performance. Second, by comprehensively integrating the local, non-local, and contextual information for LDCT image denoising, our proposed model shows the better adaptive and generalized capacity through double-blind experiments (As shown in Figure \ref{DOUBEL_1}, our proposed model has the most competitive performance under the challenge of dose mismatch between training set and test set). 
Third, how to balance the detail preservation and the level of noise removal is always a dilemma for existing deep learning-based models. However, 
our proposed model not only preserves the subtle structure and the lesion but also achieves the closest texture to NDCT (The texture can be regarded as the level of noise removal. If the level of noise removal is very high, the texture of denoising result will be very smooth). This adaptive capacity may benefit from the introduction of graph convolution, which also can be regarded as learnable Non-local Means \cite{RN413} (that has perfect adaptive ability as a non-deep-learning method). Fourth, our double-blind study is completely based on the evaluations of lesion region. We believe that this is more valuable for clinical purposes.

Compared with existing non-local-based LDCT denoising methods, our proposed model enjoys the advantages of flexible usage of non-local relationships and high computational efficiency.

In the future, various scan conditions (such as the differences of vendor, reconstruction type, and imaging parameters) will inevitably present a huge challenge for LDCT images denoising. We must be careful about this and design some methods to address it. In practice, we highly recommend  using the double-blind reader study to evaluate different denoising methods, because existing objective metrics such as Peak Signal-to-Noise Ratio (PSNR) and Structural Similarity Index Measure (SSIM)
can not fully reflect the quality of denoising results (For example, the over-smooth result usually has a high PSNR score). However, the double-blind study will take a lot of time for radiologists. We thus need some better objective metrics, especially for the region of lesion.
\section{Conclusion}
In this paper, we propose a novel graph convolutional network-based
LDCT denoising model, namely GCN-MIF, to explicitly perform
multi-information fusion for denoising purpose. Concretely, by constructing intra- and inter-slice graph, the graph convolutional
network is firstly introduced to leverage the non-local and contextual relationships among pixels. Then, the traditional CNN is still adopted for the extraction of local information. Finally, the proposed GCN-MIF model fuses all the extracted local, non-local and contextual information. Extensive experiments show the
effectiveness of proposed GCN-MIF model.

\bibliographystyle{IEEEtran}
% argument is your BibTeX string definitions and bibliography database(s)

% if have a single appendix:
%\appendix[Proof of the Zonklar Equations]
% or
%\appendix  % for no appendix heading
% do not use \section anymore after \appendix, only \section*
% is possibly needed

% use appendices with more than one appendix
% then use \section to start each appendix
% you must declare a \section before using any
% \subsection or using \label (\appendices by itself
% starts a section numbered zero.)
%

% Can use something like this to put references on a page
% by themselves when using endfloat and the captionsoff option.
\ifCLASSOPTIONcaptionsoff
  \newpage
\fi

\end{document}